\documentclass[10pt,conference]{IEEEtran}
\usepackage{cite}

\ifCLASSINFOpdf
   \usepackage[pdftex]{graphicx}
   \DeclareGraphicsExtensions{.pdf,.jpeg,.png}
\else
\fi
\usepackage[cmex10]{amsmath}
\usepackage{array}
\usepackage{url}

\usepackage{csquotes}
\usepackage{relsize}
\usepackage{adjustbox}
\usepackage{tikz}
\usepackage{pgfplots}
\usepackage{hyperref}
\usepackage{breakurl}
\pgfplotsset{compat=1.14}
\usepackage{caption}

\newcommand{\pamper}{\texttt{PaMpeR}}
\newcommand{\which}{\texttt{which\_method}}
\newcommand{\why}{\texttt{why\_method}}
\newcommand{\rank}{\texttt{rank\_method}}
\newcommand{\etal}{\textit{et al.}}
\newcommand{\comment}[1]{}

\begin{document}
%
\title{PaMpeR: Proof Method Recommendation System for Isabelle/HOL}

\author{\IEEEauthorblockN{Yutaka Nagashima}
\IEEEauthorblockA{CIIRC, Czech Technical University in Prague, Czech Republic\\
Department of Computer Science, University of Innsbruck, Austria\\
Email: yutaka.nagashima@uibk.ac.at}
\and
\IEEEauthorblockN{Yilun He}
\IEEEauthorblockA{School of Information Technologies\\
University of Sydney, Sydney, Australia\\
Email: yihe8397@uni.sydney.edu.au
}}
\maketitle

\begin{abstract}
Deciding which sub-tool to use for a given proof state requires expertise specific to each ITP.
To mitigate this problem, we present \pamper{},
a \underline{p}roof \underline{m}ethod \underline{r}ecommendation system for Isabelle/HOL.
Given a proof state, \pamper{} recommends proof methods to discharge the proof goal and
provides qualitative explanations as to why it suggests these methods.
\pamper{} generates these recommendations based on existing hand-written proof corpora,
thus transferring experienced users' expertise to new users.
Our evaluation shows that \pamper{} correctly predicts experienced users'
proof methods invocation especially when it comes to special purpose proof methods.
\end{abstract}


%
\IEEEpeerreviewmaketitle

\section{Introduction}
Do you know when to use the proof method called 
\texttt{intro\_classes} in Isabelle?
What about \texttt{uint\_arith}?
Can you tell when \verb|fast|\verb|force| tends to be more powerful than \verb|auto|?
If you are an Isabelle expert, your answer is ``\textit{Sure.}''
But if you are new to Isabelle, your answer might be 
``\textit{No. Do I have to know these Isabelle specific details?}''

Interactive theorem provers (ITPs) are forming the basis of reliable software engineering.
Klein \etal{} proved the correctness of the seL4 micro-kernel in Isabelle/HOL \cite{sel4}.
Leroy developed a certifying C compiler, CompCert, using Coq \cite{compcert}.
Kumar \etal{} built a verified compiler for a functional programming language, CakeML, in HOL4 \cite{cakeml}.
In mathematics, mathematicians are replacing their pen-and-paper proofs 
with mechanised proofs to avoid human-errors in their proofs:
Hales \etal{} mechanically proved the Kepler conjecture using HOL-light and Isabelle/HOL \cite{kepler}, whereas
Gonthier \etal{} finished the formal proofs of the four colour theorem in Coq \cite{4colour}.
In theoretical computer science,
Thiemann \etal{} formalized term rewriting system using Isabelle/HOL \cite{IsaFoR},
and Paulson proved G{\"{o}}del's incompleteness theorems using Nominal Isabelle \cite{incomplete}.

To facilitate efficient proof developments in such large scale verification projects,
modern ITPs are equipped with many sub-tools, such as proof methods and tactics.
For example, Isabelle/HOL comes with 159 proof methods defined in its standard library.
These sub-tools provide useful automation for interactive theorem proving;
however, it still requires ITP specific expertise to pick up the right proof method 
to discharge a given proof goal.

This paper presents our novel approach to proof method recommendation
and its implementation, \pamper{}.
The implementation is available at GitHub \cite{GitHub}.
Our research hypothesis is that:

\begin{displayquote}
it is possible to advise which proof methods are useful to a given proof state,
based only on the meta-information about the state and information in the standard library.
Furthermore, we can extract advice 
by applying machine learning algorithms to existing large proof corpora.
\end{displayquote} 

\noindent
The paper is organized as follows: 
Section \ref{sec:back} explains the basics of Isabelle/HOL and provides the overview of \pamper{}.
Section \ref{sec:proc} expounds how \pamper{} transforms 
the complex data structures representing proof states to simple data structures 
that are easier to handle for machine learning algorithms.
Section \ref{sec:ml} shows how our machine learning algorithm constructs regression trees 
from these simple data structures.
Section \ref{sec:reco} demonstrates how users can elicit recommendations from \pamper{}.
Section \ref{sec:eval} presents our extensive evaluation of \pamper{} 
to assess the accuracy of \pamper{}'s recommendations.
\comment{Section \ref{sec:design} justifies our design decisions for \pamper{}.}
Section \ref{sec:discussion} discusses the strengths and limitations of 
the current implementation and 
the design of a proof search tool based on \pamper{}.
Section \ref{sec:conclusion} compares our work with other attempts of applying 
machine learning and data mining to interactive theorem proving.

\section{Background and Overview of \pamper{}}\label{sec:back}

\subsection{Background} \label{sec:back1}
Isabelle/HOL is an interactive theorem prover, mostly written in Standard ML.
The consistency of Isabelle/HOL is carefully protected by isolating
its logical kernel using the module system of Standard ML.
\textit{Isabelle/Isar} \cite{isar} (\textit{Isar} for short) is a proof language used in Isabelle/HOL.
Isar provides a human-friendly interface to specify and discharge proof goals.
Isabelle users discharge proof goals by applying \textit{proof methods}, 
which are the Isar syntactic layer of LCF-style tactics. 

Each proof goal in Isabelle/HOL is stored within a \textit{proof state},
which also contains
locally bound theorems for proof methods (\textit{chained facts}) and
the background \textit{proof context} of the proof goal, 
which includes local assumptions, auxiliary definitions, and 
lemmas proved prior to the the current step.
Proof methods are in general sensitive not only to proof goals 
but also to their chained facts and background proof contexts: 
they behave differently based on information stored in proof state.
Therefore, when users decide which proof method to apply to a proof goal,
they often have to take other information in the proof state into consideration.

Isabelle comes with many Isar keywords to define new types and constants, such as
\verb|datatype|, \verb|codatatype|, \verb|primrec|, \verb|primcorec|, 
\verb|inductive|, and \verb|definition|.
For example, the \verb|fun| command is used for
general recursive definitions.

These keywords not only let users define new types or constants,
but they also automatically derive auxiliary lemmas 
relevant to the defined objects behind the user-interface and 
register them in the background proof context where each keyword is used.
For example, Nipkow \etal{} defined a function, \verb|sep|, using the \verb|fun| keyword 
in an old Isabelle tutorial \cite{NipkowPW02} as follows:

\begin{verbatim}
fun sep::"'a => 'a list => 'a list" where
"sep a [ ]      = [ ]" |
"sep a [x]      = [x]" |
"sep a (x#y#zs) = x # a # sep a (y#zs)"
\end{verbatim}

Intuitively, this function inserts the first argument between any two elements in the second argument. 
Following this definition, Isabelle automatically derives the following auxiliary lemma, 
\verb|sep.induct|, and registers it in the background proof context as well as other four automatically derived lemmas: 

\begin{verbatim}
sep.induct: (!!a. ?P a []) 
  ==> (!!a x. ?P a [x]) 
  ==> (!!a x y zs. ?P a (y # zs)
  ==> ?P a (x # y # zs)) 
  ==> ?P ?a0.0 ?a1.0
\end{verbatim}

\noindent
where variables prefixed with \verb|?| are schematic variables, 
\verb|!!| is the meta-logic universal quantifier,
\texttt{==>} is the meta-logic implication.
Isabelle also attaches unique names to these automatically derived lemmas 
following certain naming conventions hard-coded in Isabelle's source code. 
In this example, the full name of this lemma is \texttt{fun0.sep.induct}, 
which is a concatenation of the theory name (\verb|fun0|), 
the delimiter (\verb|.|), the name of the constant defined (\verb|sep|), 
followed by a hard-coded postfix (\verb|.induct|), 
which represents the kind of this derived lemma.

When users want to prove conjectures about \verb|sep|, 
they can specify their conjectures using Isar keywords such as \verb|lemma| and \verb|theorem|. 
The Isar commands, \verb|apply| and \verb|by|, 
allow users to apply proof methods to these proof goals.
In the above example, Nipkow \etal{} proved the following lemma about \verb|map| and \verb|sep| 
using the automatically derived auxiliary lemma, \verb|sep.induct|, 
as an argument to the proof method \verb|induct_tac| as following:

\begin{verbatim}
lemma 
"map f (sep x xs) = sep (f x) (map f xs)"
 apply(induct_tac x xs rule: sep.induct) 
 apply simp_all done
\end{verbatim}
\noindent
where \verb|simp_all| is a proof method that executes simplification to 
all sub-goals and \verb|done| is another Isar command used to conclude a proof attempt. 

Isabelle provides a plethora of proof methods, which serves as ammunitions 
when used by experienced Isabelle users; 
However, new Isabelle users sometimes spend hours or days trying to prove goals 
using proof methods sub-optimal to their problems 
without knowing Isabelle has already specialized methods that are optimized 
for their goals.

\comment{
Isar is an extensible languages, which let users to define new Isar commands
in Standard ML, without risking the consistency of the underlying logic.
Furthermore, Eisbach is an extension to Isar to write new proof methods
without leaving the Isar syntax.}

\subsection{Overview of \pamper{}}\label{sec:overview}
\begin{figure*}[t]
      \centerline{\includegraphics[width=150mm]{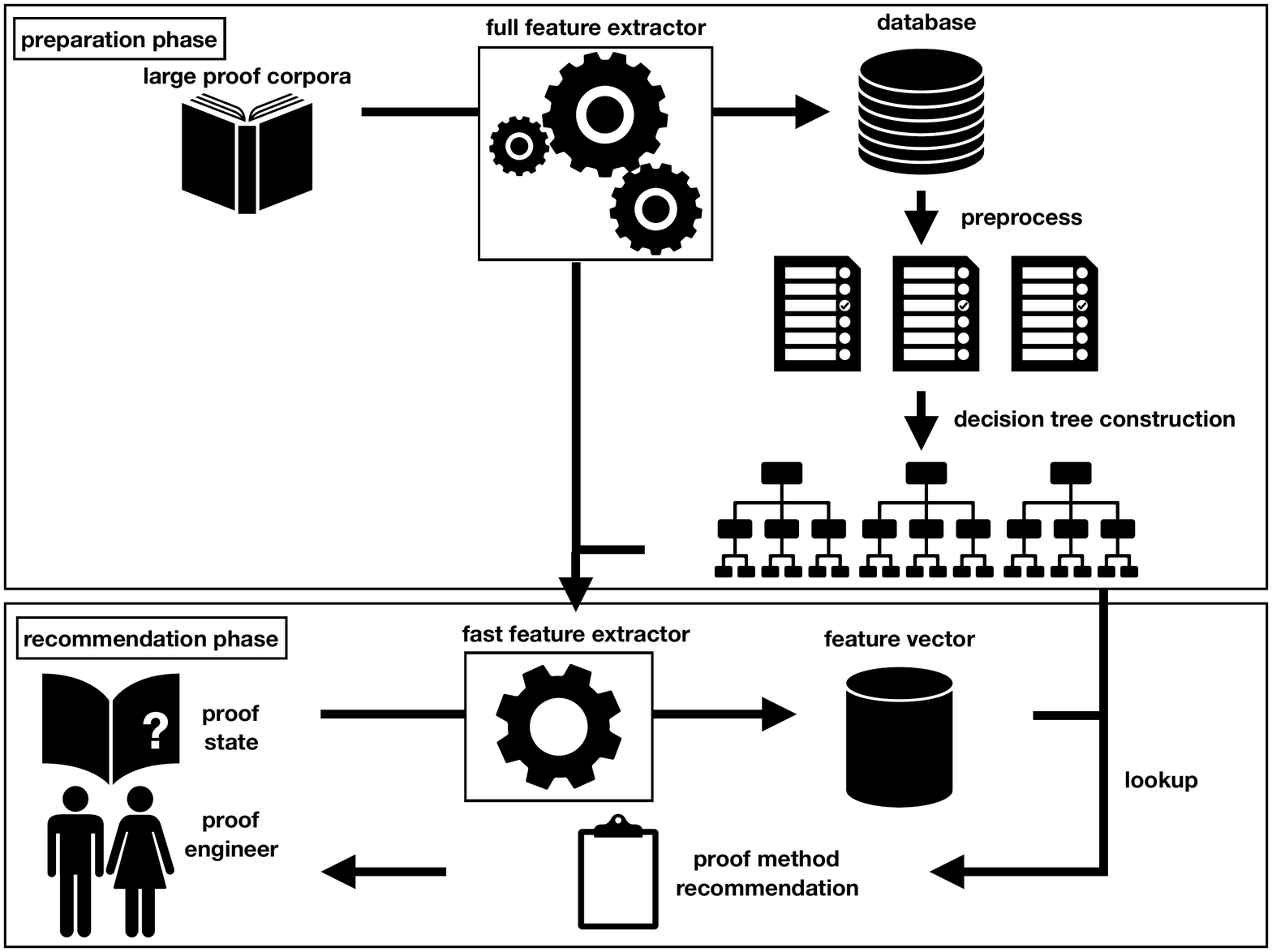}}
      \caption{Proof attempt with \pamper{}.}
      \label{fig:system}
\end{figure*}
Figure \ref{fig:system} illustrates the overview of \pamper{}.
The system consists of two phases: 
the upper half of the figure shows \pamper{}'s preparation phase,
and the lower half shows its recommendation phase.

In the preparation phase,
\pamper{}'s feature extractor converts the proof states in existing proof corpora such as
the Archive of Formal Proofs (AFP) \cite{AFP} into a database.
This database describes which proof methods have been applied to what kind of proof state,
while abstracting proof states as arrays of boolean values.
This abstraction is a many-to-one mapping:
it may map multiple distinct proof states into to the same array of boolean values.
Therefore, each array represents a group of proof states sharing certain properties.
\comment{This database is not complete: some combinations of boolean vectors are missed out,
as they do not appear in any prior proof attempt.}

\pamper{} first preprocesses this database and generates
a database for each proof method.
Then, \pamper{} applies a regression algorithm to each database and creates a regression tree for each proof method.
This regression algorithm attempts to discover combinations of 
features useful to recommend which proof method to apply.
Each tree corresponds to a certain proof method,
and each node in a tree corresponds to a group of proof states,
and the value tagged to each leaf node shows
how likely it is that the method represented by the tree is applied to these proof states
according to the proof corpora used as training sample.

For the recommendation phase, \pamper{} offers three commands, \which{}, \why{}, and \rank{}.
The \which{} command first abstracts the state into a vector of boolean values
using \pamper{}'s feature extractor.
Then, \pamper{} looks up the regression trees and 
presents its recommendations in Isabelle/jEdit's output panel.
If you wonder why \pamper{} recommends certain methods, 
for example \verb|auto|, to your proof state, 
type \verb|why_method auto|.
Then, \pamper{} tells you why it recommended \verb|auto| to the proof state
in jEdit's output panel.
If you are curious how \pamper{} ranks a certain  method, let us say \verb|intro_classes|,
type \rank{} \verb|intro_classes|.
This command shows \verb|intro_classes|'s rank given by \pamper{}
in comparison to other proof methods.
In the following, we describe these steps in detail.

\section{Processing Large Proof Corpora}\label{sec:proc}

\begin{table*}[t]
\begin{itemize}
\item Assertions about proof goals themselves.
   \begin{itemize}
   \item constants defined in Isabelle's standard library
      \begin{itemize}
         \item check if the first goal has the \verb|BNF_Def.rel_fun| constant or the \verb|Fun.map_fun| constant.
         \item check if the first goal has \verb|Orderings.ord_class.less_eq|, \verb|Orderings.ord_class.less|, or \verb|Groups.plus_class.plus|.
         \item check if the fist goal and its chained facts have \verb|Filter.eventually|
      \end{itemize}
   \item constants defined in Isabelle's standard library at certain locations in the first proof goal
      \begin{itemize}
         \item check if the outermost constant of the first goal is the meta-logic universal quantifier
         \item check if the first goal has the HOL existential quantifier but not as the outermost constant 
      \end{itemize}
   \item terms of certain types defined in Isabelle's standard library
      \begin{itemize}
         \item check if the first goal has a term of type \verb|Word.word|
         \item check if the first goal has a schematic variable
      \end{itemize}
   \item existence of constants defined in certain theory files
      \begin{itemize}
         \item check if the first goal has a constant defined in the \verb|Nat| theory
         \item check if the first goal has a constant defined in the \verb|Real| theory
         \item check if the first goal has a constant defined in the \verb|Set| theory
      \end{itemize}
   \end{itemize}
\item Assertions about the relation between proof goals and proof contexts.
   \begin{itemize}
      \item types defined with a certain Isar keyword
         \begin{itemize}
            \item check if the goal has a term of a type defined with the \verb|datatype| keyword
            \item check if the goal has a term of a type defined with the \verb|codatatype| keyword
            \item check if the goal has a term of a type defined with the \verb|record| keyword
         \end{itemize}
      \item constants defined with a certain Isar keyword
         \begin{itemize}
            \item check if the goal has a constant defined with the \verb|lift_definition| keyword
            \item check if the goal has a constant defined with the \verb|primcorec| keyword
            \item check if the goal has a constant defined with the \verb|inductive| keyword or \verb|inductive_set| keyword.
         \end{itemize}
   \end{itemize}
\end{itemize}
\caption{Selected Assertions.}
\label{list:ass}
\end{table*}

The key component of \pamper{} is its feature extractor:
the extractor converts proof goals, chained facts, and 
proof contexts into arrays of boolean values
by applying assertions to them.

\subsection{Representing a Proof State as an Array of Boolean Values}\label{sec:array}
Currently we employ 108 assertions manually written in Isabelle's 
implementation language, Standard ML,
based on our expertise in Isabelle/HOL.
Table \ref{list:ass} shows selected assertions we used in \pamper{}.
Most of these assertions fall into two categories:
assertions about proof goals themselves, and
assertions about the relation between proof goals and 
information stored in the corresponding proof context.

Note that \pamper{}'s assertions do not directly rely on any user-defined constants
because \pamper{}'s developers cannot access concrete definitions of user-defined constants when developing \pamper{}.
For example, we can check if the first proof goal 
has a constant defined in the \verb|Set.thy| file in Isabelle/HOL,
but we cannot check if that sub-goal has a constant defined in the
proof script that some user developed after we released \pamper{}.

However, by investigating how Isabelle/HOL works,
we implemented assertions that can check the meta-information of
proof goal
even without knowing their concrete specifications 
when developing \pamper{}.
For example, the lemma presented in Section \ref{sec:back1} has
a function, \verb|sep|, which was defined with the \verb|fun| keyword.
\pamper{}'s feature extractor checks if the underlying proof context contains
a lemma of name \verb|sep.elims|.
If the context has such a lemma, \pamper{} infers that 
a user defined \verb|sep| using either the \verb|fun| keyword or the \verb|function| keyword,
rather than other keywords such as \verb|primcorec| or \verb|definition|.

We wrote some assertions to reflect our own expertise in Isabelle/HOL.
One example is the assertion that checks 
if the proof goal or chained facts involve 
the constant, \verb|Filter.eventually|, defined in Isabelle's standard library.
We developed such an assertion because we knew that 
the proof method called \verb|eventually_elim| can handle 
many proof goals involving this constant.
But in some cases we were not sure which assertion 
can be useful to decide which method to use.
For example, we have assertions to check if a proof goal has constants
defined in \verb|Set.thy|, \verb|Int.thy|, or \verb|List.thy|
as these theory files define commonly used concepts in theorem proving.
But their effects to proof method selection were unclear 
until we conducted an extensive evaluation described in Section \ref{sec:eval}.

More importantly, we did not know  
numerical estimates on which assertion is more useful than others
when developing these assertions.
For instance, we guessed that the assertion to check the use of the constant \verb|Filter.eventually| 
to be useful to recommend the use of the \verb|eventually_elim| method, 
but we did not have means of comparing the accuracy of this guess with other hints prior to this project.
To obtain numerical assessments for proof method prediction,
we applied the multi-output regression algorithm described in Section \ref{sec:ml}.

\comment{
Unfortunately, not all assertions we developed turned out to be applicable
to construct a database from large proof corpora.
And we had to abandon some assertions that involve expensive operations,
as they took more memory space than we could afford
when applied to many proof states appearing in proof corpora.
Even though 
small-scale experiments indicated that
some of these abandoned assertions are useful to
predict the use of certain proof methods,
we could not even build a database from the standard library using these assertions
and decided \pamper{} serves better without them.
}


The evaluation in Section \ref{sec:eval} corroborates that 
it is possible to derive meaningful advice about proof methods.
This implies that
some parts of the expertise necessary to select appropriate proof methods 
are based on 
the meta-information about proof states or the information available 
within Isabelle's standard library,
and our assertion-based feature extractor preserves some
essence of proof states while converting them into simpler format.

\subsection{Database Extraction from Large Proof Corpora}\label{sec:database}
The first step of the preparation phase is to build a database from existing proof corpora.
We modified the proof method application commands, \verb|apply| and \verb|by|, in Isabelle and implemented a logging mechanism to build the database.
The modified \verb|apply| and \verb|by| take the following steps to generate the database:
\begin{enumerate}
\item  apply assertions to the current proof state,
\item  represent the proof state as an array of boolean values,
\item  record which method is used to that array,
\item  apply the method as the standard \verb|apply| or \verb|by| command, accordingly.
\end{enumerate}
This step requires a slight modification to the Isabelle source code 
to allow us to overwrite the definition of these command.
This way, we build its database by running the target proof scripts.

The current version of \pamper{} available at our website \cite{GitHub} is 
based on the database extracted from Isabelle's standard library
and the AFP,
but the database extraction mechanism is not specific to this library.
In case users prefer to optimise \pamper{}'s recommendation for their own proof scripts,
they can take the same approach following the instructions at our website \cite{GitHub},
even though this process tends to require significant computational resources.

This overwriting of \verb|apply| and \verb|by| is the only modification
we made to Isabelle's source code, 
and we did so only to build the database for our machine learning algorithm.
As long as users choose to use the off-the-shelf default learning results,
they can use \pamper{} without ever modifying Isabelle's source code.
In that case,
they only have to include the theory file \verb|PaMpeR/PaMpeR.thy| into their own theory file
using the Isar keyword \verb|import| just as a normal theory file to use \pamper{}.

Note that logging mechanism ignores the \verb|apply| commands 
that contain composite proof methods to avoid data pollution.
When multiple proof methods are combined within a single command, 
the naive logging approach would record proof steps 
that are backtracked to produce the final result.

One exemplary data point in an extracted database would look as the following:

\begin{verbatim}
induct, [1,0,0,1,0,0,0,0,1,0,0,1,0,...]
\end{verbatim}

\noindent
where \verb|induct| is the name of method applied to this proof state
and the $n$th element in the list shows the result of the $n$th assertion of the feature
extractor when applied to the proof state.

The default database construction from Isabelle standard library and 
the AFP took about 6021 hours 43 minutes of CPU time,
producing a database consisting of 425334 unique data points.
We used three multi-core server machines\footnote{One of them has 2 Intel(R) Xeon(R) CPUs E5-2698 v3 @ 2.30GHz with 16 cores for each and with hyperthreading, 
the other two have 2 Intel(R) Xeon(R) CPUs E5-2690 v4 @ 2.60GHz 
with 14 cores for each with hyperthreading.} 
to reduce the clock time
necessary to obtain this dataset.
Unfortunately, this database is heavily imbalanced:
some proof methods are used far more often than others.
We discuss how this imbalance influenced the quality of \pamper{}'s recommendation in Section \ref{sec:eval}.

\section{Machine Learning Databases}\label{sec:ml}

In this section, we explain the multi-output regression tree construction algorithm we implemented in Standard ML for \pamper{}.
We chose a multi-output algorithm 
because there are in general multiple valid proof methods for each proof goal,
and we chose a regression algorithm rather than classification algorithm
because we would like to provide numerical estimates about
how likely each method would be useful to a given proof goal.
We chose a regression tree construction algorithm \cite{decision_tree}
because this simple algorithm allows us to produce qualitative explanations 
as to why \pamper{} recommends certain methods
and it works well for small datasets for rarely used methods 
as shown in Section \ref{sec:eval};
However, it might be possible that 
more advanced machine learning algorithms can
result in more accurate recommendations.
The comparisons of various machine learning algorithms remain as our future work.

\subsection{Preprocess the Database}\label{sec:preproc}
We first preprocess the database generated in Section \ref{sec:database}.
This process produces a separate database for each proof method
from the raw database, which describes the use of all proof methods appearing in the target proof corpora. 

Among the class of problem transformation methods for 
multi-output regression problems,
this straightforward approach is called single-target method:
it first transforms a single multi-output problem into several
single-target problems, 
then applies a regression algorithm to each of them separately, 
then combines the results of each
regression algorithm to build a single predictor for the original
multi-output problem.

For example, 
if our preprocessor finds the example line discussed in Section \ref{sec:database},
it considers that an ideal user represented by the proof corpora
decided to use the \verb|induct| method but not other methods, 
such as \verb|auto| or \verb|coinduction|, and 
produces the following line in the database for \verb|induct|:

\begin{verbatim}
used, [1,0,0,1,0,0,0,0,1,0,0,1,0,...]
\end{verbatim}
\noindent
And the preprocessor adds the following line in the databases 
for other proof methods 
appearing in the proof corpora:

\begin{verbatim}
not, [1,0,0,1,0,0,0,0,1,0,0,1,0,...]
\end{verbatim}

Note that the resulting databases do not always represent a provably correct choice of
proof methods but conservative estimates.
In principle, there could be multiple equally valid proof methods 
for a single proof state, 
but existing proof corpora describe only one way of attacking it.
For example, Nipkow \etal{} applied the \verb|induct_tac| method to 
the lemma in Section \ref{sec:back1}, but we can prove this lemma with another method 
for mathematical induction (\verb|induction|) as follows:

\begin{verbatim}
lemma 
 "map f (sep x xs) = sep (f x) (map f xs)"
 apply(induction x xs rule: sep.induct) 
 apply simp_all done
\end{verbatim}

For this reason, this preprocessing may misjudge some methods to be inappropriate to
a proof state represented by a feature vector in some cases.
Unfortunately, exploring all the possible combinations of proof methods for each case is computational infeasible: 
some proof methods work well only when they are followed by other proof methods
or they are applied with certain arguments,
and the combination of these proof methods and arguments explodes quickly.

On the other hand, we can reasonably expect that the proof method appearing in our training sample
is the right choice to the proof state represented by the feature vector,
since Isabelle mechanically checks the proof scripts.
Furthermore, 
we built the default recommendation using Isabelle's standard
library, which was developed by experienced Isabelle developers,
and the AFP, which accepts new proofs only after peer reviews
by Isabelle experts.
This allowed us to avoid low quality proof scripts 
that Isabelle can merely process but are inappropriate.
Therefore, we consider the approximation \pamper{}'s preprocessor makes 
to be a realistic point of compromise and 
show the effectiveness of this approach
in Section \ref{sec:eval}.

\subsection{Regression Tree Construction}

After preprocessing, we apply our regression tree construction algorithm to 
each created database separately.
We implemented our tree construction algorithm from scratch in 
Standard ML for better flexibility and tool integration.

In general, the goal of the regression tree construction is to partition
the feature space described in each database into
partitions of sub-spaces that lead to
the minimal Residual Sum of Squares (RSS) 
\footnote{RSS is also known as the sum of squared residuals (SSR).}
while avoiding over-fitting.
Intuitively, RSS denotes the discrepancy between 
the data and estimation based on a model.
The RSS in our problem is defined as follows:
\begin{equation}
RSS = 
\mathlarger{\mathlarger{\sum}}_{j=1}^{J}
\mathlarger{\mathlarger{\sum}}_{i \in R_j}
(used_i - \widehat{used}_{R_j})^2
\end{equation}
\noindent
where $R_j$ stands for the $j$th sub-space, 
to which certain data points (represented as lines in database) belong.
The value of $used_i$ is
$1.0$ if the data point represented by the subscript 
$i$ says the method was applied to the feature vector,
and it is $0.0$ if the data point represented by the subscript $i$ says otherwise.
$\widehat{used}_{R_j}$ is the average value of $used$ among the data points
pertaining to the sub-space $R_j$.

\comment{
The construction algorithm computes the residual sum of squares (RSS) to decide
which feature to use to split the predictor space described in each database.
The definition of RSS is as follows:
The splitting algorithm uses the residual sum of squares (RSS) to decide which
feature to focus.}

Computing the RSS for every possible partition of the database under consideration 
is computational infeasible.
Therefore,
\pamper{}'s tree construction takes a top-down, greedy approach,
called \textit{recursive binary splitting} \cite{I2SL}.

In recursive binary splitting,
we start constructing the regression tree from the root node,
which corresponds to the entire dataset for a given method.
First, we select a feature in such a way we can 
achieve the greatest reduction in RSS at this particular step.
We find such feature by computing the reduction of the RSS by each feature by one level.
For each feature, we split the database into two sub-spaces,
$R_{used}(j)$ and $R_{not}(j)$ as follows:
\begin{equation}
\begin{split}
&R_{used}(j) = \{used | used_j = 1.0\} \text{ and }\\
&R_{not}(j) = \{used | used_j = 0.0\}
\end{split}
\end{equation}
\noindent
where $j$ stands for the number representing each feature.
Then, for each feature represented by $j$, we compute the following value:

\begin{equation}
\begin{split}
&\mathlarger{\mathlarger{\sum}}_{i: x_i \in R_{used}(j)}
  (used_i - \widehat{used}_{R_{used}(j)})^2 + \\
&\mathlarger{\mathlarger{\sum}}_{i: x_i \in R_{not}(j)}
  (used_i - \widehat{used}_{R_{not}(j)})^2
\end{split}
\end{equation}\label{min_rss}
\noindent
and choose the feature $j$ that minimizes this value.

Second, we repeat this partition procedure to each emerging sub-node of the regression tree
under construction until the depth of tree hits our pre-defined upper limit. 

After reaching the maximum depth,
we compute the average value of $used(j)$ 
in the corresponding sub-space $R$ for each leaf node.
We consider this value as the expectation that the method is useful to
proof states abstracted to the combination of feature values to that leaf node.

\pamper{} records these regression trees in a text file, \verb|regression_trees.txt|, so that
users can avoid the computationally intensive data extraction and regression tree construction processes
unless they want to optimize the learning results based on their own proof corpora.

Note that if we add more assertions to our feature extractor in future,
the complexity of this algorithm increases linearly with the number of assertions 
given a fixed depth of regression tree,
since the partition only takes the best step at each level instead of exploring
all the combinations of partitions.

\section{Recommendation Phase}\label{sec:reco}

Once finishing building regression trees for each proof method appeared in the 
given proof corpora, one can extract recommendations from \pamper{}.
When imported to users' theory file,
\pamper{} automatically reads these trees using the \verb|read_regression_trees| command in \verb|PaMpeR/PaMpeR.thy|.

\pamper{} provides three new commands to provide two kinds of information: 
the \which{} command tells which proof methods are likely to be useful for a given proof state;
the \why{} command takes a name of proof method and tells why \pamper{} would recommend the proof method for the proof state;
and the \rank{} command shows the rank of a given method to the proof state in comparison to other proof methods.
In the following, we explain how these two commands produce recommendations from the regression trees produced in the preparation phase.

\subsection{Faster Feature Extractor}\label{sec:read_tree}
Before applying the machine learning algorithm, we were not sure
which assertion produces valuable features,
but after applying the machine learning algorithm,
we can judge which assertions are not useful,
by checking which features are used to branch each regression tree.

The \texttt{build\_fast\_feature\_extractor} command in \verb|PaMpeR/PaMpeR.thy|
constructs a faster feature extractor from the regression trees 
built in the preparation phase and the full feature extractor
to reduce the waiting time of \pamper{}'s users.
It builds the faster feature extractor by removing assertions that do not result in a branch in the regression trees.

\subsection{The \which{} command.}\label{sec:which}
When users invoke the \which{} command, 
\pamper{} applies the faster feature extractor to convert 
the ongoing proof state into a feature vector, which consists of those features that are
deemed to be important to make a recommendation.
The speed of this faster feature vector depends on both the regression trees and 
what each proof state contains.
As a rule of thumb, if the proof goal has less terms, 
it tends to spend less time.

Then, \pamper{} looks up the corresponding node in each regression tree
and decides the expectation that the method is the right choice for the 
proof state represented by the feature vector.
\pamper{} computes this value for each proof method 
it encountered in the training proof corpora,
by looking up a node in each regression tree.
Finally, \pamper{} compares these expectations and shows
the 15 most promising proof methods with their expectations 
in Isabelle/jEdit's output panel. 
In the on-going example from Section \ref{sec:back1}, 
a user can know which method to use by typing the \which{} command as follows:

\begin{verbatim}
lemma 
 "map f (sep x xs) = sep (f x) (map f xs)"
 which_method
\end{verbatim}
\noindent
Then, \pamper{} shows the following message in the output panel for the top 15 methods
\footnote{Note that we truncated the message due to the space restriction here.}:

\begin{verbatim}
Promising methods for this proof goal are: 
  simp with expectation of 0.4119
  auto with expectation of 0.1593 
  rule with expectation of 0.0874 
  induction with expectation of 0.06137 
  metis with expectation of 0.05260 ... 
\end{verbatim}
\noindent
Attentive readers might have noticed that \pamper{}'s recommendations are not 
identical to the model answer provided by Nipkow \etal{}
This, however, does not immediately mean \pamper{}'s recommendation is not valuable:
in fact, \pamper{} recommended the \verb|induction| method at the fourth place
out of 239 proof methods, 
and \verb|induction| is also a valid method for this proof goal 
as discussed in Section \ref{sec:preproc}.

\subsection{The \why{} command.}\label{sec:why}
Our rather straightforward machine learning algorithm
makes \pamper{}'s recommendation \textit{explainable}.
If you wonder why \pamper{} recommends a certain method, for example \verb|case_tac|,
to your proof goal, type \verb|why_method case_tac| in the proof script.
\pamper{} first checks features used to evaluate the expectation for the 
method and their feature values.
Second, \pamper{} shows qualitative explanations tagged to both these features and their values
in jEdit's output.
If you wonder why \pamper{} recommended \verb|induction| in the above example, type the following:
\begin{verbatim}
lemma 
 "map f (sep x xs) = sep (f x) (map f xs)"
 why_method induction
\end{verbatim}
\noindent
Then, you will see this message in jEdit's output panel:
\begin{verbatim}
Because it is not true that the context 
has locally defined assumptions. 
Because the underlying proof context has 
a recursive simplification rule related to 
a constant appearing in the first subgoal.
\end{verbatim}
\indent
The first reason corresponds to the first branching at the root node
in the regression tree for the \verb|induction| method, and the second reason
corresponds to the second branching in the tree.
In this case, \pamper{} found that the proof goal involves the constant,
\verb|sep|, and the underlying proof context contains a simplification rule,
\verb|sep.simps(3)|, which involves a recursive call of \verb|sep| as following:
\begin{verbatim}
sep.simp(3):
  sep ?a (?x # ?y # ?zs) 
= ?x # ?a # sep ?a (?y # ?zs)
\end{verbatim}

\subsection{The \rank{} command.}\label{sec:rank}
Sometimes users already have a guess as to 
which proof method would be useful to their proof state, 
but they want to know how \pamper{} ranks the proof method in mind. 
Continuing with the above example, 
if you want to know how \pamper{} ranks \verb|conduction| for this proof state,
type the following:
\noindent
\begin{verbatim}
lemma
"map f (sep x xs) = sep (f x) (map f xs)"
 rank_method coinduction
\end{verbatim}
\noindent 
Then, \pamper{} warns you:

\begin{verbatim}coinduction 123 out of 239\end{verbatim}
\noindent
indicating that \pamper{} does not consider 
\verb|coinduction| to be the right choice for this proof goal,
before you waste your time on emerging sub-goals appearing after applying \verb|coinduction|.



\section{Evaluation}\label{sec:eval}

\setlength{\tabcolsep}{3pt}
\begin{table*}[!ht]
\caption{Evaluation of \pamper{} on 15 most frequently used proof methods.}
\label{table:top15}
\begin{center}
\begin{tabular}{l r r r r r r r r r r r r r r r r r r r r r r}
\hline\noalign{\smallskip}
proof method & training & \% & evaluation & \% & 1 & 2 & 3 & 4 & 5 & 6 & 7 & 8 & 9 & 10 & 11 & 12 & 13 & 14 & 15\\
\hline
\noalign{\smallskip}
\verb|simp| & 102441 & 26.8 & 11385 & 26.8 & 58 & 98 & 99 & 99 & 100 & 100 & 100 & 100 & 100 & 100 & 100 & 100 & 100 & 100 & 100\\
\verb|auto| & 85097 & 22.2 & 9527 & 22.4 & 60 & 94 & 98 & 99 & 100 & 100 & 100 & 100 & 100 & 100 & 100 & 100 & 100 & 100 & 100\\
\verb|rule| & 38856 & 10.2 & 4150 & 9.8 & 3 & 15 & 86 & 99 & 100 & 100 & 100 & 100 & 100 & 100 & 100 & 100 & 100 & 100 & 100\\
\verb|blast| & 23814 & 6.2 & 2590 & 6.1 & 0 & 26 & 26 & 35 & 84 & 95 & 99 & 100 & 100 & 100 & 100 & 100 & 100 & 100 & 100\\
\verb|metis| & 19771 & 5.2 & 2149 & 5.1 & 0 & 0 & 13 & 72 & 84 & 89 & 93 & 96 & 98 & 99 & 100 & 100 & 100 & 100 & 100\\
\verb|fastforce| & 9477 & 2.5 & 1093 & 2.6 & 0 & 0 & 0 & 0 & 5 & 54 & 70 & 81 & 89 & 93 & 96 & 96 & 97 & 98 & 98\\
\verb|force| & 6232 & 1.6 & 708 & 1.7 & 0 & 0 & 0 & 0 & 1 & 9 & 22 & 32 & 40 & 51 & 66 & 77 & 84 & 89 & 94\\
\verb|clarsimp| & 5984 & 1.6 & 628 & 1.5 & 0 & 12 & 14 & 14 & 20 & 29 & 39 & 49 & 54 & 57 & 62 & 64 & 66 & 67 & 73\\
\verb|cases| & 5842 & 1.5 & 689 & 1.6 & 0 & 0 & 1 & 16 & 16 & 20 & 34 & 54 & 70 & 80 & 86 & 91 & 93 & 95 & 96\\
\verb|erule| & 5732 & 1.5 & 707 & 1.7 & 0 & 0 & 15 & 38 & 44 & 53 & 64 & 70 & 76 & 82 & 85 & 87 & 91 & 92 & 93\\
\verb|subst| & 5655 & 1.5 & 619 & 1.5 & 0 & 0 & 19 & 19 & 19 & 20 & 22 & 28 & 45 & 58 & 69 & 77 & 82 & 86 & 90\\
\verb|rule_tac| & 5342 & 1.4 & 631 & 1.5 & 0 & 14 & 32 & 34 & 44 & 45 & 46 & 47 & 50 & 51 & 52 & 52 & 53 & 57 & 63\\
\verb|intro| & 4988 & 1.3 & 619 & 1.5 & 0 & 0 & 5 & 18 & 24 & 39 & 46 & 47 & 48 & 48 & 49 & 57 & 69 & 77 & 84\\
\verb|simp_all| & 4982 & 1.3 & 568 & 1.3 & 0 & 0 & 0 & 1 & 3 & 6 & 15 & 21 & 26 & 33 & 45 & 60 & 70 & 78 & 83\\
\verb|induct| & 4884 & 1.3 & 568 & 1.3 & 0 & 0 & 0 & 1 & 27 & 45 & 49 & 50 & 50 & 51 & 56 & 62 & 71 & 77 & 79\\
\hline
\end{tabular}
\end{center}
\end{table*}

We conducted a cross-validation to assess the accuracy of \pamper{}'s \which{} command.
For this evaluation, 
we used Isabelle's standard library and the AFP as follows:
First, we extracted a database from these proof corpora.
This database consists of 425334 data points.
Second, we randomly chose 10\% of data points in this database to create the evaluation dataset.
Third, we built regression trees from the remaining 90\%.
There is no overlap between the evaluation dataset and training dataset.
Then, we applied regression trees to each each data point in the evaluation dataset and
counted how often \pamper{}'s recommendation coincides with the proof methods chosen by
human proof authors.



Since there are often multiple equally valid proof methods for each proof state,
it is only reasonable to expect that
\which{} should be able to 
recommend the proof method used in the evaluation dataset as one of the most important methods
for each proof method invocation.
Therefore, for each proof method,
we measured how often each proof method used in the evaluation dataset
appears among the top $n$ methods in \pamper{}'s recommendations.


Table \ref{table:top15} shows the results for the 15 proof methods 
that are most frequently used in the training data in the descending order.

For example, the top row for \verb|simp| should be interpreted as following:
The \verb|simp| method was used 102441 times in the training data.
This amounts to 26.8\% of all proof method invocations in the training data
that are recorded by \pamper{}.
In the evaluation dataset, \verb|simp| was used 11385 times, 
which amounts to 26.8\% of 
proof method invocations in the evaluation dataset 
that are recorded by \pamper{}.
For 58\% out of 11385 \verb|simp| invocations in the evaluation dataset, 
\pamper{} predicted that \verb|simp| is the most promising method 
for the corresponding proof states.
For 98\% out of 11385 \verb|simp| invocations in the evaluation dataset,
\pamper{} recommended that \verb|simp| is either the most promising method
or the second most promising method for the corresponding proof states.

Note that the numbers presented in this table are not  
the success rates of \pamper{}'s recommendation but its conservative estimates.
Assume \pamper{} recommends \verb|simp| as the most promising method 
and \verb|auto| as the second most promising method to a proof goal,
say \verb|pg|, in the evaluation dataset, 
but the human proof author of \verb|pg| chose to 
apply \verb|auto| to this proof goal. 
This does not immediately mean that \pamper{} failed to recommend \verb|auto|
in the first place, because
both \verb|simp| and \verb|auto| might be equally suitable for \verb|pg|.
Therefore, the 58\% for \verb|simp| mentioned above should be interpreted as follows:
\pamper{}'s recommendation coincides with the choice of 
experienced Isabelle user 
for 58\% of times where human engineers applied \verb|simp|
when \pamper{} is allowed to recommend only one proof method,
but the real success rate of \pamper{}'s recommendation
can be higher than 58\% for these cases.
To avoid the confusion with \textit{success rate},
we introduce the term, \textit{coincidence rate}, for this measure.
Table \ref{table:standard1}, \ref{table:standard2} and, \ref{table:user}
in Appendix provide the complete list of the evaluation results.

The overall results of this evaluation are as follows:
\pamper{} learnt 239 proof methods from Isabelle's standard library and the AFP:
160 of them are defined within Isabelle's standard library, 
and the others are user-defined proof methods, which are 
specified in the AFP entries.

Out of the 239 proof methods \pamper{} learnt from the training dataset,
171 proof methods appeared in the evaluation dataset.
Out of these 171 proof methods within the evaluation dataset,
133 methods are defined in Isabelle's standard library,
and 38 methods were defined by the AFP authors.

The distribution of proof method usage is heavily imbalanced.
The three most frequently used proof methods 
(\verb|simp|, \verb|auto|, and \verb|rule|) account for 59.1\% of all
data points in the training dataset, 
and the ten most frequently used methods account for 79.2\% in the training dataset.
Similarly in the evaluation dataset,
the top three methods account for 58.9\%, and the top ten methods for 79.1\%.

Fig. 2 illustrates this imbalance, 
in which the horizontal axis represents the rank of method usage for
a proof method
and the vertical axis stands for the number of methods invocations
for that proof method.
For instance, the square located at the top-left corner denotes that
the most frequently used proof method in the training dataset (\verb|simp|)
is used 102441 times. And the circle located at $(6, 1093)$ denotes
that the sixth most frequently used method in the evaluation dataset 
(\verb|fastforce|) is used 1093 times in the evaluation dataset.
With the use of logarithmic scale on the vertical axis,
this figure presents the serious imbalance of proof method invocations
occurring in Isabelle's standard library and the AFP.

Fig. 3 summarises the overall performance of \pamper{}.
In this figure the horizontal axis represents the number of 
proof methods \pamper{} is allowed to recommend (15 by default), whereas
the vertical axis represents the number of proof methods, for which
\pamper{} achieves certain coincidence rates.

For example, the square at $(3, 23)$ means that
\pamper{} can achieve 50\% of coincidence rate for 23 methods
if \pamper{} is allowed to recommend three most promising methods.
Similarly, \pamper{} achieves 50\% of coincidence rate 
for 58 methods when recommending 10 methods and
for 72 methods when recommending 15 methods.

The number of methods \pamper{} achieved 
the four coincident rates (25\%, 50\%, 75\%, and 90\%),
reached a plateau when \pamper{} is allowed to recommend
about 60 proof methods.

Overall, \pamper{}'s recommendations 
tend to coincide with human engineers' choice when Isabelle has only
one method that is suitable for the proof goal at hand, whereas
\pamper{}'s recommendations tend to differ from human engineers' choice 
when there are multiple equally valid proof methods for the same goal.
For example, \pamper{}'s coincidence rates are low for less commonly used
general-purpose methods, such as \verb|safe|, \verb|clarimp|, 
\verb|best|, \verb|bestsimp|
because multiple general purpose proof methods can 
often handle the same proof goal equally well.

A careful observation at the raw evaluation results provided in the Appendix
reveal that 
\pamper{} provides valuable recommendations when proof states are best handled 
by special purpose proof methods, 
such as \verb|unfold_locales|, \verb|transfer|, \verb|eventually_elim|, \verb|standard|, and so on.

\pamper{}'s regression tree construction does not 
severely suffer from the imbalance among proof method invocation, 
even though class imbalances often cause problems 
in other domains such as
fraud detection and medical diagnosis \cite{DBLP:journals/tkde/HeG09}.
The complete evaluation results in Appendix show that
\pamper{} achieved 50\% of coincidence rate for 34 proof methods
that appear less than 0.1\% of times in the training dataset.

The reason the imbalance did not cause serious problems to \pamper{} is that
some of these rarely used methods are specialised proof methods,
for which we can write assertions that can abstract 
the essence of the problem very well. 
Another reason is the fact that commonly used proof methods
tend to hold up each other's share,
since they address similar problems, 
lowering expectations for commonly used general purpose methods 
where both specialised methods and general purpose methods
can discharge proof goals.

On the other hand, \pamper{} did not produce valuable recommendations to some 
special purpose proof methods, such as \verb|normalization| and \verb|vector|, 
for which we did not manage to develop assertions that capture
the properties shared by the proof goals that these
methods can handle well.
Writing suitable assertions for these remain as our future work.

Some of the proof methods appearing in our evaluation dataset are clearly
outside the scope of \pamper{}. 
For example, \verb|tactic|, \verb|cartouche|, \verb|ml_tactic|,
\verb|rotate_tac| do not have much semantic meaning:
\verb|tactic| is simply an interface between Isabelle's source code language,
Standard ML, and Isabelle's proof language, Isar,
whereas \verb|rotate_tac| simply rotates the order of premises when
a proof goal has multiple premises. 
Another good example of proof methods outside the scope of \pamper{} 
is the \verb|my_simp| method.
This method was defined in the standard library to test
the domain specific language, \textit{Eisbach},
for writing new proof methods:
\verb|my_simp| is simply a synonym of \verb|simp| and 
nobody is expected to use \verb|my_simp|.
Predicting such methods is not a very meaningful task for \pamper{}.


To our surprise, Table \ref{table:user} shows that
\pamper{}'s recommendation achieved 50\% of coincidence rate
for 12 methods out of 38 user-defined proof methods defined outside Isabelle's standard library appearing in the evaluation dataset when
\pamper{} is allowed to provide 15 most promising proof methods,
even though \pamper{}'s developers did not know anything about these proof methods at the time of development.
This suggests that 
one does not need to know the problem specific information about 
proof goals to predict the use of some user-defined proof methods.
For example, 
\pamper{} achieves 100\% of coincidence rate for \verb|sepref| when
allowed to recommend only four methods,
by checking if the first sub-goal has a schematic variable 
and if the first sub-goal has variables of type record.

\begin{figure}[t]
\begin{tikzpicture}
\begin{axis}[
    title={Fig.2: Method usage in large proof corpora.},
    ylabel={Number of method invocation.},
    xlabel={Nth most commonly used proof  method.},
    xmin=1, xmax=169,
    ymin=1, ymax=103000,
    xtick={0,20,40,60,80,100,120,140,160},
    legend pos=north east,
    ymode=log,
    log ticks with fixed point,
    only marks
]

\addplot[
    color=black,
    mark=square,
    ]
    coordinates {
    (1, 102441) (2, 85097) (3, 38856) (4, 23814) (5, 19771) (6, 9477) (7, 6232) (8, 5984) (9, 5842) (10, 5732) (11, 5655) (12, 5342) (13, 4988) (14, 4982) (15, 4884) (16, 3347) (17, 3078) (18, 3020) (19, 2981) (20, 2861) (21, 2856) (22, 2653) (23, 2620) (24, 1855) (25, 1839) (26, 1661) (27, 1592) (28, 1491) (29, 1353) (30, 1275) (31, 1251) (32, 1127) (33, 1003) (34, 998) (35, 982) (36, 735) (37, 681) (38, 622) (39, 621) (40, 587) (41, 550) (42, 534) (43, 529) (44, 492) (45, 475) (46, 461) (47, 361) (48, 288) (49, 282) (50, 263) (51, 257) (52, 235) (53, 235) (54, 206) (55, 194) (56, 193) (57, 193) (58, 173) (59, 167) (60, 162) (61, 158) (62, 152) (63, 126) (64, 120) (65, 118) (66, 113) (67, 112) (68, 108) (69, 103) (70, 102) (71, 94) (72, 91) (73, 90) (74, 90) (75, 89) (76, 86) (77, 80) (78, 78) (79, 77) (80, 73) (81, 73) (82, 70) (83, 67) (84, 66) (85, 66) (86, 64) (87, 63) (88, 58) (89, 58) (90, 55) (91, 49) (92, 45) (93, 44) (94, 44) (95, 43) (96, 43) (97, 42) (98, 41) (99, 40) (100, 39) (101, 37) (102, 36) (103, 35) (104, 35) (105, 35) (106, 33) (107, 32) (108, 31) (109, 29) (110, 29) (111, 23) (112, 23) (113, 23) (114, 23) (115, 22) (116, 22) (117, 21) (118, 21) (119, 21) (120, 19) (121, 19) (122, 18) (123, 18) (124, 17) (125, 17) (126, 16) (127, 15) (128, 15) (129, 15) (130, 15) (131, 15) (132, 14) (133, 13) (134, 11) (135, 10) (136, 10) (137, 9) (138, 8) (139, 8) (140, 8) (141, 7) (142, 7) (143, 6) (144, 6) (145, 5) (146, 5) (147, 5) (148, 5) (149, 5) (150, 4) (151, 4) (152, 4) (153, 4) (154, 3) (155, 3) (156, 3) (157, 3) (158, 3) (159, 2) (160, 2) (161, 2) (162, 2) (163, 2) (164, 2) (165, 2) (166, 1) (167, 1) (168, 1) (169, 1)
    };
    ]
    \addlegendentry{in training dataset}

\addplot[
    color=black,
    mark=*,
    mark options={fill=white},
    ]
    coordinates {
    (1, 11385) (2, 9527) (3, 4150) (4, 2590) (5, 2149) (6, 1093) (7, 708) (8, 707) (9, 689) (10, 631) (11, 628) (12, 619) (13, 619) (14, 568) (15, 568) (16, 364) (17, 362) (18, 356) (19, 343) (20, 332) (21, 299) (22, 293) (23, 279) (24, 193) (25, 186) (26, 185) (27, 184) (28, 166) (29, 153) (30, 139) (31, 133) (32, 127) (33, 102) (34, 100) (35, 97) (36, 76) (37, 71) (38, 71) (39, 65) (40, 65) (41, 59) (42, 57) (43, 57) (44, 55) (45, 50) (46, 48) (47, 37) (48, 32) (49, 31) (50, 30) (51, 28) (52, 26) (53, 26) (54, 22) (55, 20) (56, 18) (57, 18) (58, 17) (59, 17) (60, 17) (61, 16) (62, 16) (63, 15) (64, 15) (65, 14) (66, 14) (67, 14) (68, 13) (69, 12) (70, 12) (71, 12) (72, 12) (73, 12) (74, 12) (75, 12) (76, 10) (77, 10) (78, 9) (79, 8) (80, 7) (81, 7) (82, 7) (83, 7) (84, 7) (85, 7) (86, 6) (87, 6) (88, 6) (89, 6) (90, 6) (91, 6) (92, 6) (93, 6) (94, 5) (95, 5) (96, 5) (97, 5) (98, 5) (99, 4) (100, 4) (101, 4) (102, 4) (103, 4) (104, 4) (105, 4) (106, 4) (107, 4) (108, 4) (109, 4) (110, 3) (111, 3) (112, 3) (113, 3) (114, 3) (115, 3) (116, 2) (117, 2) (118, 2) (119, 2) (120, 2) (121, 2) (122, 2) (123, 2) (124, 2) (125, 2) (126, 2) (127, 2) (128, 2) (129, 2) (130, 2) (131, 2) (132, 2) (133, 2) (134, 2) (135, 2) (136, 2) (137, 2) (138, 1) (139, 1) (140, 1) (141, 1) (142, 1) (143, 1) (144, 1) (145, 1) (146, 1) (147, 1) (148, 1) (149, 1) (150, 1) (151, 1) (152, 1) (153, 1) (154, 1) (155, 1) (156, 1) (157, 1) (158, 1) (159, 1) (160, 1) (161, 1) (162, 1) (163, 1) (164, 1) (165, 1) (166, 1) (167, 1) (168, 1) (169, 1)
    };
    ]
    \addlegendentry{in evaluation dataset}

\end{axis}
\end{tikzpicture}
\label{method_usage}
\end{figure}

\begin{figure}[t]
\begin{tikzpicture}
\begin{axis}[
    title={Fig. 3: Coincidence rate for \pamper{}.},
    xlabel={Number of methods \pamper{} recommends.},
    ylabel={Methods above four coincidence rates.},
    xmin=1, xmax=169,
    ymin=1, ymax=169,
    xtick={1,20,40,60,80,100,120,140,160},
    ytick={1,20,40,60,80,100,120,140,160},
    legend pos=south east,
    grid style=dashed,
    only marks
]

\addplot[
    color=black,
    mark=triangle,
    ]
    coordinates {
    (1, 14)  (2, 18)  (3, 29)  (4, 40)  (5, 43)  (6, 48)  (7, 60)  (8, 70)  (9, 71)  (10, 75)  (11, 78)  (12, 81)  (13, 84)  (14, 86)  (15, 86)  (16, 90)  (17, 93)  (18, 93)  (19, 97)  (20, 98)  (21, 103)  (22, 105)  (23, 108)  (24, 109)  (25, 109)  (26, 113)  (27, 113)  (28, 115)  (29, 117)  (30, 120)  (31, 123)  (32, 123)  (33, 126)  (34, 127)  (35, 129)  (36, 131)  (37, 131)  (38, 133)  (39, 134)  (40, 135)  (41, 138)  (42, 138)  (43, 139)  (44, 139)  (45, 141)  (46, 142)  (47, 145)  (48, 147)  (49, 148)  (50, 149)  (51, 150)  (52, 150)  (53, 151)  (54, 153)  (55, 153)  (56, 155)  (57, 156)  (58, 156)  (59, 156)  (60, 158)  (61, 159)  (62, 159)  (63, 159)  (64, 159)  (65, 159)  (66, 159)  (67, 159)  (68, 159)  (69, 159)  (70, 160)  (71, 160)  (72, 162)  (73, 162)  (74, 162)  (75, 162)  (76, 162)  (77, 163)  (78, 163)  (79, 163)  (80, 163)  (81, 163)  (82, 163)  (83, 163)  (84, 163)  (85, 163)  (86, 163)  (87, 163)  (88, 163)  (89, 163)  (90, 163)  (91, 163)  (92, 163)  (93, 163)  (94, 163)  (95, 163)  (96, 163)  (97, 163)  (98, 163)  (99, 163)  (100, 163)  (101, 164)  (102, 164)  (103, 164)  (104, 164)  (105, 164)  (106, 164)  (107, 164)  (108, 164)  (109, 164)  (110, 164)  (111, 164)  (112, 164)  (113, 164)  (114, 164)  (115, 164)  (116, 164)  (117, 164)  (118, 164)  (119, 164)  (120, 164)  (121, 164)  (122, 164)  (123, 164)  (124, 164)  (125, 164)  (126, 164)  (127, 164)  (128, 164)  (129, 164)  (130, 164)  (131, 164)  (132, 164)  (133, 164)  (134, 164)  (135, 164)  (136, 165)  (137, 165)  (138, 165)  (139, 165)  (140, 165)  (141, 165)  (142, 165)  (143, 165)  (144, 165)  (145, 165)  (146, 165)  (147, 165)  (148, 165)  (149, 165)  (150, 166)  (151, 166)  (152, 166)  (153, 166)  (154, 166)  (155, 166)  (156, 166)  (157, 166)  (158, 166)  (159, 166)  (160, 166)  (161, 166)  (162, 166)  (163, 166)  (164, 166)  (165, 166)  (166, 166)  (167, 166)  (168, 166)  (169, 166)
    };
    ]
    \addlegendentry{more than $25\%$}
    
\addplot[
    color=black,
    mark=square,
    ]
    coordinates {
     (1, 13)  (2, 16)  (3, 23)  (4, 28)  (5, 29)  (6, 34)  (7, 40)  (8, 45)  (9, 53)  (10, 58)  (11, 63)  (12, 67)  (13, 70)  (14, 70)  (15, 72)  (16, 75)  (17, 76)  (18, 81)  (19, 85)  (20, 86)  (21, 90)  (22, 92)  (23, 93)  (24, 94)  (25, 96)  (26, 98)  (27, 98)  (28, 100)  (29, 105)  (30, 108)  (31, 112)  (32, 114)  (33, 117)  (34, 120)  (35, 121)  (36, 123)  (37, 123)  (38, 126)  (39, 127)  (40, 129)  (41, 130)  (42, 132)  (43, 134)  (44, 134)  (45, 136)  (46, 136)  (47, 140)  (48, 142)  (49, 144)  (50, 145)  (51, 146)  (52, 146)  (53, 146)  (54, 149)  (55, 149)  (56, 151)  (57, 151)  (58, 153)  (59, 154)  (60, 156)  (61, 158)  (62, 158)  (63, 158)  (64, 158)  (65, 158)  (66, 158)  (67, 158)  (68, 158)  (69, 158)  (70, 159)  (71, 159)  (72, 160)  (73, 160)  (74, 160)  (75, 160)  (76, 160)  (77, 161)  (78, 162)  (79, 162)  (80, 162)  (81, 163)  (82, 163)  (83, 163)  (84, 163)  (85, 163)  (86, 163)  (87, 163)  (88, 163)  (89, 163)  (90, 163)  (91, 163)  (92, 163)  (93, 163)  (94, 163)  (95, 163)  (96, 163)  (97, 163)  (98, 163)  (99, 163)  (100, 163)  (101, 164)  (102, 164)  (103, 164)  (104, 164)  (105, 164)  (106, 164)  (107, 164)  (108, 164)  (109, 164)  (110, 164)  (111, 164)  (112, 164)  (113, 164)  (114, 164)  (115, 164)  (116, 164)  (117, 164)  (118, 164)  (119, 164)  (120, 164)  (121, 164)  (122, 164)  (123, 164)  (124, 164)  (125, 164)  (126, 164)  (127, 164)  (128, 164)  (129, 164)  (130, 164)  (131, 164)  (132, 164)  (133, 164)  (134, 164)  (135, 164)  (136, 165)  (137, 165)  (138, 165)  (139, 165)  (140, 165)  (141, 165)  (142, 165)  (143, 165)  (144, 165)  (145, 165)  (146, 165)  (147, 165)  (148, 165)  (149, 165)  (150, 166)  (151, 166)  (152, 166)  (153, 166)  (154, 166)  (155, 166)  (156, 166)  (157, 166)  (158, 166)  (159, 166)  (160, 166)  (161, 166)  (162, 166)  (163, 166)  (164, 166)  (165, 166)  (166, 166)  (167, 166)  (168, 166)  (169, 166)
    };
    ]
    \addlegendentry{more than $50\%$}

\addplot[
    color=black,
    mark=*,
    mark options={fill=white},
    ]
    coordinates {
     (1, 6)  (2, 10)  (3, 15)  (4, 18)  (5, 22)  (6, 24)  (7, 25)  (8, 28)  (9, 34)  (10, 36)  (11, 37)  (12, 40)  (13, 44)  (14, 48)  (15, 50)  (16, 54)  (17, 57)  (18, 59)  (19, 62)  (20, 65)  (21, 68)  (22, 69)  (23, 70)  (24, 75)  (25, 79)  (26, 79)  (27, 83)  (28, 84)  (29, 87)  (30, 90)  (31, 93)  (32, 97)  (33, 99)  (34, 102)  (35, 106)  (36, 108)  (37, 110)  (38, 112)  (39, 114)  (40, 117)  (41, 118)  (42, 120)  (43, 123)  (44, 125)  (45, 127)  (46, 128)  (47, 129)  (48, 129)  (49, 134)  (50, 135)  (51, 137)  (52, 138)  (53, 138)  (54, 141)  (55, 142)  (56, 144)  (57, 145)  (58, 146)  (59, 146)  (60, 149)  (61, 150)  (62, 151)  (63, 151)  (64, 151)  (65, 151)  (66, 151)  (67, 151)  (68, 152)  (69, 153)  (70, 153)  (71, 153)  (72, 154)  (73, 154)  (74, 154)  (75, 154)  (76, 154)  (77, 155)  (78, 155)  (79, 155)  (80, 156)  (81, 157)  (82, 157)  (83, 158)  (84, 158)  (85, 160)  (86, 160)  (87, 160)  (88, 160)  (89, 160)  (90, 160)  (91, 160)  (92, 160)  (93, 160)  (94, 160)  (95, 160)  (96, 160)  (97, 160)  (98, 160)  (99, 160)  (100, 160)  (101, 161)  (102, 162)  (103, 162)  (104, 162)  (105, 162)  (106, 162)  (107, 162)  (108, 162)  (109, 162)  (110, 162)  (111, 162)  (112, 162)  (113, 162)  (114, 162)  (115, 162)  (116, 162)  (117, 162)  (118, 162)  (119, 162)  (120, 162)  (121, 162)  (122, 162)  (123, 162)  (124, 162)  (125, 162)  (126, 162)  (127, 162)  (128, 162)  (129, 162)  (130, 162)  (131, 162)  (132, 162)  (133, 162)  (134, 162)  (135, 162)  (136, 163)  (137, 163)  (138, 163)  (139, 163)  (140, 163)  (141, 163)  (142, 163)  (143, 163)  (144, 163)  (145, 163)  (146, 163)  (147, 163)  (148, 163)  (149, 163)  (150, 164)  (151, 164)  (152, 164)  (153, 164)  (154, 164)  (155, 164)  (156, 164)  (157, 164)  (158, 164)  (159, 164)  (160, 164)  (161, 164)  (162, 164)  (163, 164)  (164, 164)  (165, 164)  (166, 164)  (167, 164)  (168, 164)  (169, 164)
    };
    ]
    \addlegendentry{more than $75\%$}

\addplot[
    color=black,
    mark=x,
    ]
    coordinates {
     (1, 5)  (2, 9)  (3, 12)  (4, 15)  (5, 15)  (6, 17)  (7, 20)  (8, 22)  (9, 25)  (10, 28)  (11, 28)  (12, 29)  (13, 31)  (14, 31)  (15, 34)  (16, 34)  (17, 37)  (18, 40)  (19, 44)  (20, 47)  (21, 49)  (22, 52)  (23, 54)  (24, 58)  (25, 59)  (26, 61)  (27, 63)  (28, 64)  (29, 67)  (30, 69)  (31, 74)  (32, 75)  (33, 79)  (34, 85)  (35, 88)  (36, 91)  (37, 92)  (38, 96)  (39, 99)  (40, 101)  (41, 104)  (42, 105)  (43, 107)  (44, 108)  (45, 115)  (46, 117)  (47, 119)  (48, 119)  (49, 121)  (50, 122)  (51, 124)  (52, 125)  (53, 126)  (54, 129)  (55, 132)  (56, 134)  (57, 135)  (58, 135)  (59, 136)  (60, 138)  (61, 140)  (62, 142)  (63, 142)  (64, 143)  (65, 143)  (66, 143)  (67, 143)  (68, 143)  (69, 145)  (70, 145)  (71, 146)  (72, 147)  (73, 147)  (74, 148)  (75, 149)  (76, 150)  (77, 151)  (78, 151)  (79, 151)  (80, 152)  (81, 153)  (82, 153)  (83, 154)  (84, 154)  (85, 155)  (86, 155)  (87, 156)  (88, 156)  (89, 156)  (90, 156)  (91, 156)  (92, 156)  (93, 156)  (94, 157)  (95, 157)  (96, 157)  (97, 157)  (98, 157)  (99, 157)  (100, 157)  (101, 158)  (102, 159)  (103, 159)  (104, 159)  (105, 159)  (106, 159)  (107, 159)  (108, 159)  (109, 159)  (110, 159)  (111, 159)  (112, 159)  (113, 159)  (114, 159)  (115, 159)  (116, 159)  (117, 159)  (118, 159)  (119, 159)  (120, 161)  (121, 161)  (122, 161)  (123, 161)  (124, 161)  (125, 161)  (126, 161)  (127, 161)  (128, 161)  (129, 161)  (130, 161)  (131, 161)  (132, 161)  (133, 161)  (134, 161)  (135, 161)  (136, 162)  (137, 162)  (138, 162)  (139, 162)  (140, 162)  (141, 162)  (142, 162)  (143, 162)  (144, 162)  (145, 162)  (146, 162)  (147, 162)  (148, 162)  (149, 162)  (150, 163)  (151, 163)  (152, 163)  (153, 163)  (154, 163)  (155, 163)  (156, 163)  (157, 163)  (158, 163)  (159, 163)  (160, 163)  (161, 163)  (162, 163)  (163, 163)  (164, 163)  (165, 163)  (166, 163)  (167, 163)  (168, 163)  (169, 163)
    };
    ]
    \addlegendentry{more than $90\%$}

\end{axis}
\end{tikzpicture}
\label{graph}
\end{figure}

\section{Discussion and Future Work}\label{sec:discussion}

Prior to \pamper{}, Isabelle had the \verb|print_methods| command,
which merely lists the proof methods defined in the corresponding proof context 
in alphabetical order ignoring the properties of the proof goal at hand.
Therefore, new Isabelle/HOL users have to go through various 
documentations and the archive of mailing lists
to learn how to prove lemmas in Isabelle/HOL independently.

Choosing the right methods was a difficult task for new ITP users especially 
when they should choose special-purpose proof methods, since new
users tend not to know even the existence of those rarely used proof methods. 
Some proof methods are strongly related to certain definitional mechanisms in Isabelle.
Therefore, when Isabelle experts use such definitional mechanisms,
they can often guess which proof methods they should use later.
But this is not an easy task for new users.
And this is becoming truer nowadays,
since large scale theorem proving projects are slowly becoming popular and 
new ITP users often have to take over proof scripts developed by others
and they also have to discharge proof goals specified by others.
\pamper{} addressed this problem by systematically transferring 
experienced users’ knowledge to less experienced users.
We plan to keep improving \pamper{} by incorporating other Isabelle users intuitions as assertions. 

Our manually written feature extractor may seem to be naive 
compared to the recent success in machine learning research:
in some problem domains, such as image recognition and the game of Go,
deep neural networks extract features of the subject matters via expensive training.
Indeed, others have applied deep neural networks to theorem proving, but without much success \cite{deepmath, dngps}.

The two major problems of automatic feature extraction for theorem proving is 
the lack of enormous database needed to train deep neural networks and 
the expressive nature of the underlying language, i.e. logic. 
The second problem, the expressive nature of logic, contributes to the first problem:
self-respecting proof engineers tend to replace multiple similar propositions 
with one proposition from which one can easily conclude similar propositions,
aiming at a succinct presentation of the underlying concept.

What is worse,
when working with modern ITPs,
it is often not enough to reason about a proof goal,
but one also has to take its proof context into consideration.
A proof context usually contains numerous auxiliary lemmas and nested definitions, and each of them is a syntax tree, 
making the effective automatic feature extraction harder.

Furthermore, whenever a proof author defines a new constant or prove a new lemma,
Isabelle/HOL changes the underlying proof context, 
which affects how one should attack proof goals defined within this proof context.
And proof authors do add new definitions because they use ITPs as
specification tools as well as tools for theorem proving.
Some of these changes are minor modifications to proof states that do not severely affect how to attack proof goals in the following proof scripts,
but in general changing proof contexts results in, sometimes unexpected, problems.

For this reason, even though the ITP community has large proof corpora,
we are essentially dealing with different problems in each line of each proof corpus.
For example, even the AFP has 396 articles consisting of more than 100,000 lemmas,
only 4 articles are used by more than 10 articles in the AFP,
indicating that many proof authors work on their own specifications, 
creating new problems.
This results in an important difference that lies 
between theorem proving in an expressive logic 
and other machine learning domains, such as image recognition
where one can collects numerous instances of similar objects.


We addressed this problem with human-machine cooperation, 
the philosophy that underpins ITPs.
Even though it is hard to extract features automatically,
experienced ITP users know that they can discharge many proof goals with shallow reasoning.
We encoded experienced Isabelle users' expertise as assertions to simulate their shallow reasoning.
Since these assertions are carefully hand-written in Isabelle/ML,
they can extract features of proof states (including proof goal, chained facts, and its context)
despite the above mentioned problems.


Currently \pamper{} recommends only which methods to use
and shows why it suggests that method.
This is enough for many special purpose methods which do not take parameters.
For other methods, such as \verb|induct|, 
it is often indispensable to pass the correct parameters to guide methods.
If you prefer to know which arguments to pass to the proof method \pamper{} recommends,
we would invite you to use \verb|PSL| \cite{psl},
the \underline{p}roof \underline{s}trategy \underline{l}anguage for Isabelle/HOL,
which attempts to find the right combination of arguments
through an iterative deepening depth first search
based on rough ideas about which method to use.
If you want to have those rough ideas, use \pamper{}.

Moreover, none of \pamper{}'s assertions takes the sequence of proof method applications into account:
even though they can check the information contained in the background proof context,
the parse-then-consume style of Isabelle/Isar makes it difficult for \pamper{} to trace 
which methods have been applied to reach the current proof state.

We envision a more powerful proof automation tool backed by \pamper{} and \verb|PSL|.
We use \pamper{}'s recommendation to navigate the search of \verb|PSL|,
changing \verb|PSL|'s evaluation strategy from the IDDFS
to the best-first search.
\verb|PSL| provides a mechanism to generate variants of proof methods with different combinations of parameters
to find the right combination of parameters for a given goal through a search.
\verb|PSL|'s automatic removal of backtracked proof steps eliminates the data pollution problem discussed in Section \ref{sec:database}.
\verb|PSL|'s framework to write history-sensitive proof methods 
allows us to write history-sensitive assertions,
so that \pamper{} can take the sequence of proof methods into account.

\section{Conclusion and Related Work}\label{sec:conclusion}

We presented the design and implementation of \pamper{}.
In the preparation phase,
\pamper{} learns which method to use from existing proof corpora using regression tree
construction algorithm.
In the recommendation phase,
\pamper{} recommends which proof methods to use to a given proof goal
and explains why it suggests that method.
Our evaluation showed that \pamper{} tends to provide valuable recommendations 
especially for specialised proof methods,
which new Isabelle users tend not to be aware of.
We also identified problems that arise when applying machine learning to proof method recommendation
and proposed our solution to them.

\paragraph*{Related Work}
ML4PG \cite{ml4pg} extends a proof editor, Proof General, to collect proof statistics about shapes of goals,
sequence of applied tactics, and proof tree structures.
It also clusters the gathered data using machine learning algorithms in MATLAB and Weka
and provides proof hints during proof developments.
Based on learning, ML4PG lists similar proof goals proved so far,
from which users can infer how to attack the proof goal at hand,
while \pamper{} directly works on proof methods.
Compared to ML4PG, \pamper{}'s feature extractor is implemented within Isabelle/ML,
which made it possible to investigate not only proof goals themselves 
but also their surrounding proof context.

Gauthier \etal{} developed TacticToe for HOL4.
It selects proved lemmas similar to the current proof goal using premise selection
and applies tactics used to these similar goals to discharge the current proof goal.
Compared to TacticToe,
the abstraction via assertions allows \pamper{} to provide valuable recommendations
even when similar goals do not exist in the problem domain.

Several people applied machine learning techniques to improve the so-called Hammer-style tools.
For Isabelle/HOL, both MePo \cite{mepo} and MaSh \cite{mesh} decreased the quantity of facts passed to the automatic provers while increasing their quality to improve Sledgehammer’s performance. Their approaches attempt to choose facts that are likely to be useful to the given proof goal, while \pamper{} suggests proof methods that are likely to be useful to the goal.

MePo judges the relevance of facts by checking the occurrence of symbols appearing in proof goals and available facts, 
while MaSh computes the relevance using sparse naive Bayes and k Nearest Neighbours. 
They detect similarities between proof goals and available facts by checking mostly formalization-specific information and only two piece of meta information, 
while \pamper{} discards most of problem specific information and focus on meta information of proof goals: the choice of relevant fact is a problem specific question, while the choice of proof method largely depends on which Isabelle's subsystem is used to specify a proof goal.


The original version of MaSh was using machine learning libraries in Python, 
and Blanchette \etal{} ported them from Python to Standard ML for better efficiency and reliability. 
Similarly, an early version of \pamper{} was also using a Python library \cite{scikit} until we implemented the regression tree construction algorithm in Standard ML for better tool integration and flexibility.
Both MaSh and \pamper{} record learning results in persistent states outside the main memory, so that users can preserve the learning results even after shutting down Isabelle.

Blanchette \etal{} analysed the AFP, looking at sizes and dependencies
for theory files \cite{mining_jasmin}.
Matichuk \etal{} investigated the seL4 proofs and two articles in the AFP
to find the relationship between the size of statement and the size of proof \cite{mining_daniel}.
None of them analysed the occurrence of proof methods in their target proof 
corpora nor developed a recommendation system based on their results.
Moreover, \pamper{}'s database construction is more active compared to their work:
it applies 108 hand-written assertions to analyse the properties of
not only each proof goal but also the relationship between each goal and
its background context and chained facts.

\newpage

\section*{Appendix}

Table \ref{table:standard1} shows the results for proof methods 
that are 
defined in the standard library and are
frequently used in the training data in the descending order.
Table \ref{table:standard2} shows the results for proof methods 
that are 
defined in the standard library and are
less frequently used in the training data in the descending order.
Table \ref{table:user} shows the results for proof methods 
that are locally defined in the AFP entries by Isabelle users.

For example, the top row for \verb|simp| in Table \ref{table:standard1}
should be interpreted as following:
The \verb|simp| method was used 102441 times in the training data.
This amounts to 26.8\% of all proof method invocations in the training data
that are recorded by \pamper{}.
In the evaluation dataset, \verb|simp| was used 11385 times, 
which amounts to 26.8\% of 
proof method invocations in the evaluation dataset 
that are recorded by \pamper{}.
For 58\% out of 11385 \verb|simp| invocations in the evaluation dataset, 
\pamper{} predicted that \verb|simp| is the most promising method 
for the corresponding proof states.
For 98\% out of 11385 \verb|simp| invocations in the evaluation dataset,
\pamper{} recommended that \verb|simp| is either the most promising method
or the second most promising method for the corresponding proof states.

Note that the numbers presented in this table are not  
the success rates of \pamper{}'s recommendation but its conservative estimates.
Assume \pamper{} recommends \verb|simp| as the most promising method 
and \verb|auto| as the second most promising method to a proof goal,
say \verb|pg|, in the evaluation dataset, 
but the human proof author of \verb|pg| chose to 
apply \verb|auto| to this proof goal. 
This does not immediately mean that \pamper{} failed to recommend \verb|auto|
in the first place, because
both \verb|simp| and \verb|auto| might be equally suitable for \verb|pg|.
Therefore, the 58\% for \verb|simp| mentioned above should be interpreted as follows:
\pamper{}'s recommendation coincides with the choice of 
experienced Isabelle user 
for 58\% of times where human engineers applied \verb|simp|
when \pamper{} is allowed to recommend only one proof method,
but the real success rate of \pamper{}'s recommendation
can be higher than 58\% for these cases.
In Section \ref{sec:eval},
we introduced the term, \textit{coincidence rate}, for this measure
to avoid the confusion with \textit{success rate}.

The last figure in the Appendix 
is a screenshot of Isabelle/HOL with \pamper{},
which shows the seamless integration of \pamper{} to Isabelle's
default proof editor, Isabelle/jEdit.

\setlength{\tabcolsep}{3pt}
\begin{table*}[hp]
\caption{Evaluation of \pamper{} on proof methods defined in the standard library 1.}
\label{table:standard1}
\begin{center}
\begin{tabular}{l r r r r r r r r r r r r r r r r r r r r r r}
\hline\noalign{\smallskip}
proof method & training & \% & evaluation & \% & 1 & 2 & 3 & 4 & 5 & 6 & 7 & 8 & 9 & 10 & 11 & 12 & 13 & 14 & 15\\
\hline
\noalign{\smallskip}
\verb|simp| & 102441 & 26.8 & 11385 & 26.8 & 58 & 98 & 99 & 99 & 100 & 100 & 100 & 100 & 100 & 100 & 100 & 100 & 100 & 100 & 100\\
\verb|auto| & 85097 & 22.2 & 9527 & 22.4 & 60 & 94 & 98 & 99 & 100 & 100 & 100 & 100 & 100 & 100 & 100 & 100 & 100 & 100 & 100\\
\verb|rule| & 38856 & 10.2 & 4150 & 9.8 & 3 & 15 & 86 & 99 & 100 & 100 & 100 & 100 & 100 & 100 & 100 & 100 & 100 & 100 & 100\\
\verb|blast| & 23814 & 6.2 & 2590 & 6.1 & 0 & 26 & 26 & 35 & 84 & 95 & 99 & 100 & 100 & 100 & 100 & 100 & 100 & 100 & 100\\
\verb|metis| & 19771 & 5.2 & 2149 & 5.1 & 0 & 0 & 13 & 72 & 84 & 89 & 93 & 96 & 98 & 99 & 100 & 100 & 100 & 100 & 100\\
\verb|fastforce| & 9477 & 2.5 & 1093 & 2.6 & 0 & 0 & 0 & 0 & 5 & 54 & 70 & 81 & 89 & 93 & 96 & 96 & 97 & 98 & 98\\
\verb|force| & 6232 & 1.6 & 708 & 1.7 & 0 & 0 & 0 & 0 & 1 & 9 & 22 & 32 & 40 & 51 & 66 & 77 & 84 & 89 & 94\\
\verb|clarsimp| & 5984 & 1.6 & 628 & 1.5 & 0 & 12 & 14 & 14 & 20 & 29 & 39 & 49 & 54 & 57 & 62 & 64 & 66 & 67 & 73\\
\verb|cases| & 5842 & 1.5 & 689 & 1.6 & 0 & 0 & 1 & 16 & 16 & 20 & 34 & 54 & 70 & 80 & 86 & 91 & 93 & 95 & 96\\
\verb|erule| & 5732 & 1.5 & 707 & 1.7 & 0 & 0 & 15 & 38 & 44 & 53 & 64 & 70 & 76 & 82 & 85 & 87 & 91 & 92 & 93\\
\verb|subst| & 5655 & 1.5 & 619 & 1.5 & 0 & 0 & 19 & 19 & 19 & 20 & 22 & 28 & 45 & 58 & 69 & 77 & 82 & 86 & 90\\
\verb|rule_tac| & 5342 & 1.4 & 631 & 1.5 & 0 & 14 & 32 & 34 & 44 & 45 & 46 & 47 & 50 & 51 & 52 & 52 & 53 & 57 & 63\\
\verb|intro| & 4988 & 1.3 & 619 & 1.5 & 0 & 0 & 5 & 18 & 24 & 39 & 46 & 47 & 48 & 48 & 49 & 57 & 69 & 77 & 84\\
\verb|simp_all| & 4982 & 1.3 & 568 & 1.3 & 0 & 0 & 0 & 1 & 3 & 6 & 15 & 21 & 26 & 33 & 45 & 60 & 70 & 78 & 83\\
\verb|induct| & 4884 & 1.3 & 568 & 1.3 & 0 & 0 & 0 & 1 & 27 & 45 & 49 & 50 & 50 & 51 & 56 & 62 & 71 & 77 & 79\\
\hline
\noalign{\smallskip}
\verb|case_tac| & 3347 & 0.9 & 362 & 0.9 & 0 & 0 & 0 & 0 & 10 & 24 & 35 & 43 & 49 & 54 & 58 & 62 & 66 & 70 & 77\\
\verb|fast| & 3078 & 0.8 & 364 & 0.9 & 0 & 0 & 0 & 0 & 0 & 0 & 0 & 7 & 22 & 45 & 57 & 62 & 64 & 67 & 70\\
\verb|drule| & 3020 & 0.8 & 343 & 0.8 & 0 & 0 & 0 & 0 & 7 & 19 & 24 & 25 & 29 & 35 & 45 & 55 & 68 & 79 & 90\\
\verb|subgoal_tac| & 2981 & 0.8 & 332 & 0.8 & 0 & 0 & 0 & 0 & 2 & 11 & 28 & 44 & 61 & 70 & 73 & 76 & 78 & 79 & 82\\
\verb|fact| & 2861 & 0.7 & 356 & 0.8 & 0 & 0 & 30 & 30 & 30 & 30 & 33 & 35 & 39 & 50 & 60 & 68 & 75 & 80 & 81\\
\verb|unfold| & 2856 & 0.7 & 293 & 0.7 & 0 & 0 & 0 & 0 & 0 & 12 & 30 & 45 & 52 & 58 & 60 & 61 & 61 & 61 & 61\\
\verb|-| & 2653 & 0.7 & 279 & 0.7 & 0 & 0 & 0 & 0 & 0 & 2 & 15 & 32 & 36 & 37 & 38 & 42 & 51 & 66 & 80\\
\verb|drule_tac| & 2620 & 0.7 & 299 & 0.7 & 0 & 7 & 23 & 38 & 38 & 38 & 38 & 39 & 40 & 41 & 43 & 48 & 52 & 64 & 71\\
\verb|assumption| & 1855 & 0.5 & 184 & 0.4 & 0 & 0 & 39 & 45 & 48 & 48 & 49 & 50 & 50 & 50 & 50 & 52 & 52 & 55 & 61\\
\verb|induction| & 1839 & 0.5 & 185 & 0.4 & 0 & 0 & 0 & 37 & 40 & 40 & 40 & 40 & 40 & 40 & 40 & 40 & 40 & 41 & 45\\
\verb|transfer| & 1661 & 0.4 & 186 & 0.4 & 0 & 0 & 51 & 51 & 51 & 51 & 51 & 51 & 51 & 51 & 52 & 53 & 56 & 59 & 60\\
\verb|frule| & 1592 & 0.4 & 166 & 0.4 & 0 & 0 & 0 & 0 & 1 & 5 & 16 & 29 & 39 & 42 & 42 & 42 & 42 & 42 & 43\\
\verb|erule_tac| & 1491 & 0.4 & 193 & 0.5 & 0 & 0 & 0 & 1 & 5 & 17 & 32 & 46 & 52 & 58 & 62 & 63 & 63 & 63 & 63\\
\verb|clarify| & 1353 & 0.4 & 153 & 0.4 & 0 & 0 & 0 & 0 & 2 & 9 & 18 & 31 & 33 & 33 & 38 & 39 & 40 & 44 & 48\\
\verb|meson| & 1275 & 0.3 & 139 & 0.3 & 2 & 2 & 2 & 2 & 2 & 2 & 2 & 2 & 2 & 2 & 2 & 2 & 3 & 8 & 9\\
\verb|rename_tac| & 1251 & 0.3 & 133 & 0.3 & 0 & 0 & 0 & 0 & 0 & 2 & 5 & 13 & 23 & 30 & 42 & 53 & 62 & 68 & 72\\
\verb|unfold_locales| & 1127 & 0.3 & 97 & 0.2 & 0 & 8 & 50 & 50 & 50 & 50 & 50 & 50 & 50 & 50 & 50 & 50 & 50 & 50 & 50\\
\verb|cut_tac| & 1003 & 0.3 & 102 & 0.2 & 0 & 0 & 0 & 0 & 0 & 0 & 0 & 0 & 0 & 5 & 8 & 17 & 26 & 28 & 36\\
\verb|arith| & 998 & 0.3 & 127 & 0.3 & 0 & 0 & 0 & 0 & 0 & 0 & 18 & 27 & 42 & 48 & 50 & 51 & 53 & 53 & 54\\
\verb|safe| & 982 & 0.3 & 100 & 0.2 & 0 & 0 & 0 & 0 & 0 & 0 & 0 & 3 & 6 & 7 & 7 & 7 & 11 & 13 & 14\\
\verb|frule_tac| & 735 & 0.2 & 71 & 0.2 & 0 & 0 & 0 & 0 & 1 & 3 & 3 & 3 & 3 & 3 & 3 & 3 & 3 & 3 & 3\\
\verb|elim| & 681 & 0.2 & 71 & 0.2 & 0 & 0 & 0 & 0 & 0 & 0 & 0 & 0 & 0 & 0 & 0 & 0 & 0 & 0 & 0\\
\verb|induct_tac| & 622 & 0.2 & 76 & 0.2 & 0 & 0 & 0 & 14 & 14 & 14 & 14 & 14 & 14 & 14 & 14 & 14 & 14 & 14 & 14\\
\verb|eval| & 621 & 0.2 & 65 & 0.2 & 0 & 0 & 0 & 8 & 48 & 62 & 65 & 65 & 65 & 65 & 65 & 65 & 65 & 65 & 65\\
\verb|iprover| & 587 & 0.2 & 65 & 0.2 & 0 & 0 & 0 & 0 & 0 & 11 & 11 & 11 & 11 & 11 & 11 & 15 & 20 & 28 & 31\\
\verb|linarith| & 550 & 0.1 & 57 & 0.1 & 0 & 0 & 4 & 4 & 4 & 9 & 9 & 9 & 10 & 10 & 10 & 10 & 10 & 10 & 10\\
\verb|standard| & 534 & 0.1 & 59 & 0.1 & 56 & 56 & 56 & 56 & 56 & 56 & 59 & 66 & 70 & 70 & 71 & 71 & 71 & 71 & 71\\
\verb|presburger| & 529 & 0.1 & 55 & 0.1 & 0 & 0 & 0 & 0 & 0 & 6 & 11 & 20 & 20 & 20 & 20 & 20 & 20 & 20 & 20\\
\verb|thin_tac| & 492 & 0.1 & 50 & 0.1 & 0 & 0 & 0 & 0 & 0 & 0 & 0 & 0 & 0 & 0 & 0 & 0 & 2 & 12 & 24\\
\verb|insert| & 475 & 0.1 & 37 & 0.1 & 0 & 0 & 0 & 0 & 0 & 0 & 0 & 0 & 0 & 0 & 0 & 0 & 0 & 0 & 0\\
\verb|tactic| & 461 & 0.1 & 57 & 0.1 & 0 & 0 & 16 & 16 & 16 & 16 & 16 & 16 & 16 & 16 & 16 & 16 & 16 & 16 & 16\\
\verb|nominal_induct| & 361 & 0.1 & 48 & 0.1 & 0 & 0 & 0 & 19 & 23 & 23 & 23 & 23 & 23 & 23 & 23 & 23 & 25 & 29 & 54\\
\verb|rotate_tac| & 288 & 0.1 & 30 & 0.1 & 0 & 0 & 0 & 0 & 0 & 0 & 0 & 3 & 3 & 3 & 3 & 17 & 17 & 17 & 17\\
\verb|perm_simp| & 282 & 0.1 & 32 & 0.1 & 0 & 0 & 0 & 0 & 0 & 0 & 3 & 12 & 12 & 16 & 19 & 19 & 47 & 47 & 56\\
\verb|eventually_elim| & 263 & 0.1 & 28 & 0.1 & 0 & 0 & 86 & 86 & 86 & 86 & 86 & 100 & 100 & 100 & 100 & 100 & 100 & 100 & 100\\
\verb|transfer_prover| & 257 & 0.1 & 26 & 0.1 & 50 & 50 & 50 & 73 & 77 & 77 & 77 & 77 & 77 & 77 & 77 & 77 & 77 & 77 & 77\\
\verb|generate_fresh| & 235 & 0.1 & 22 & 0.1 & 0 & 0 & 0 & 0 & 0 & 0 & 0 & 0 & 18 & 27 & 36 & 41 & 41 & 41 & 41\\
\verb|contradiction| & 235 & 0.1 & 18 & 0.0 & 0 & 0 & 0 & 0 & 0 & 0 & 0 & 0 & 0 & 0 & 78 & 78 & 78 & 78 & 78\\
\verb|vcg| & 206 & 0.1 & 26 & 0.1 & 88 & 92 & 92 & 92 & 92 & 92 & 92 & 92 & 92 & 92 & 92 & 92 & 92 & 92 & 92\\
\verb|smt| & 194 & 0.1 & 20 & 0.0 & 0 & 0 & 0 & 0 & 0 & 0 & 0 & 0 & 0 & 0 & 0 & 0 & 0 & 0 & 5\\
\verb|intro_classes| & 193 & 0.1 & 14 & 0.0 & 0 & 100 & 100 & 100 & 100 & 100 & 100 & 100 & 100 & 100 & 100 & 100 & 100 & 100 & 100\\
\verb|fold| & 162 & 0.0 & 15 & 0.0 & 0 & 0 & 0 & 7 & 7 & 7 & 7 & 7 & 7 & 7 & 7 & 7 & 7 & 7 & 7\\
\verb|rewrite| & 158 & 0.0 & 17 & 0.0 & 0 & 0 & 6 & 6 & 6 & 6 & 6 & 6 & 6 & 6 & 6 & 6 & 6 & 6 & 6\\
\verb|fresh_fun_simp| & 152 & 0.0 & 15 & 0.0 & 0 & 0 & 0 & 0 & 0 & 40 & 47 & 47 & 73 & 73 & 73 & 73 & 73 & 73 & 73\\
\verb|measurable| & 126 & 0.0 & 16 & 0.0 & 0 & 0 & 0 & 0 & 0 & 0 & 0 & 62 & 75 & 94 & 100 & 100 & 100 & 100 & 100\\
\verb|pat_completeness| & 120 & 0.0 & 12 & 0.0 & 92 & 92 & 92 & 92 & 92 & 92 & 92 & 92 & 92 & 100 & 100 & 100 & 100 & 100 & 100\\
\verb|ind_cases| & 118 & 0.0 & 13 & 0.0 & 0 & 0 & 0 & 0 & 0 & 0 & 0 & 0 & 0 & 0 & 0 & 0 & 0 & 0 & 0\\
\verb|field| & 112 & 0.0 & 14 & 0.0 & 0 & 0 & 36 & 36 & 36 & 36 & 36 & 36 & 36 & 36 & 36 & 36 & 36 & 36 & 43\\
\verb|sos| & 103 & 0.0 & 14 & 0.0 & 50 & 50 & 50 & 50 & 50 & 50 & 50 & 50 & 50 & 50 & 50 & 50 & 50 & 50 & 50\\
\verb|relation| & 102 & 0.0 & 12 & 0.0 & 92 & 92 & 92 & 92 & 92 & 92 & 92 & 92 & 92 & 92 & 92 & 92 & 92 & 92 & 92\\
\hline
\end{tabular}
\end{center}
\end{table*}

\begin{table*}[hp]
\caption{Evaluation of \pamper{} on proof methods defined in the standard library 2.}
\label{table:standard2}
\begin{center}
\begin{tabular}{l r r r r r r r r r r r r r r r r r r r r r r}
\hline\noalign{\smallskip}
proof method & training & \% & evaluation & \% & 1 & 2 & 3 & 4 & 5 & 6 & 7 & 8 & 9 & 10 & 11 & 12 & 13 & 14 & 15\\
\hline
\noalign{\smallskip}
\verb|descending| & 94 & 0.0 & 18 & 0.0 & 0 & 0 & 0 & 0 & 0 & 0 & 6 & 33 & 61 & 61 & 61 & 61 & 61 & 61 & 61\\
\verb|normalization| & 91 & 0.0 & 12 & 0.0 & 25 & 25 & 25 & 25 & 25 & 25 & 25 & 25 & 25 & 25 & 25 & 25 & 25 & 25 & 25\\
\verb|fixrec_simp| & 90 & 0.0 & 9 & 0.0 & 0 & 89 & 89 & 89 & 89 & 89 & 89 & 89 & 89 & 89 & 89 & 89 & 89 & 89 & 89\\
\verb|coinduct| & 90 & 0.0 & 6 & 0.0 & 0 & 0 & 0 & 0 & 0 & 0 & 0 & 0 & 0 & 0 & 67 & 67 & 83 & 83 & 83\\
\verb|lexicographic_order| & 89 & 0.0 & 12 & 0.0 & 67 & 67 & 67 & 92 & 92 & 92 & 92 & 92 & 92 & 92 & 92 & 92 & 92 & 92 & 92\\
\verb|argo| & 78 & 0.0 & 16 & 0.0 & 0 & 0 & 0 & 0 & 0 & 0 & 0 & 0 & 0 & 6 & 6 & 6 & 12 & 12 & 12\\
\verb|ring| & 73 & 0.0 & 7 & 0.0 & 0 & 0 & 0 & 43 & 43 & 43 & 43 & 43 & 43 & 43 & 43 & 43 & 43 & 43 & 43\\
\verb|atomize_elim| & 67 & 0.0 & 4 & 0.0 & 0 & 0 & 0 & 0 & 0 & 0 & 0 & 0 & 0 & 0 & 0 & 0 & 0 & 0 & 0\\
\verb|coinduction| & 66 & 0.0 & 12 & 0.0 & 0 & 0 & 0 & 25 & 33 & 33 & 33 & 33 & 33 & 33 & 33 & 33 & 33 & 33 & 33\\
\verb|best| & 64 & 0.0 & 4 & 0.0 & 0 & 0 & 0 & 0 & 0 & 0 & 0 & 0 & 0 & 0 & 0 & 0 & 0 & 0 & 0\\
\verb|this| & 63 & 0.0 & 10 & 0.0 & 0 & 0 & 0 & 0 & 0 & 0 & 0 & 0 & 0 & 0 & 0 & 0 & 0 & 0 & 0\\
\verb|hypsubst_thin| & 58 & 0.0 & 6 & 0.0 & 0 & 0 & 0 & 0 & 0 & 0 & 0 & 0 & 0 & 0 & 0 & 0 & 0 & 0 & 0\\
\verb|spy_analz| & 55 & 0.0 & 6 & 0.0 & 0 & 0 & 0 & 0 & 0 & 0 & 0 & 0 & 0 & 0 & 0 & 0 & 0 & 0 & 0\\
\verb|pair| & 49 & 0.0 & 7 & 0.0 & 14 & 14 & 14 & 14 & 14 & 14 & 14 & 14 & 14 & 14 & 57 & 57 & 57 & 71 & 71\\
\verb|ferrack| & 45 & 0.0 & 2 & 0.0 & 0 & 0 & 0 & 0 & 0 & 0 & 50 & 100 & 100 & 100 & 100 & 100 & 100 & 100 & 100\\
\verb|disjE_tac| & 44 & 0.0 & 2 & 0.0 & 0 & 0 & 0 & 0 & 0 & 0 & 0 & 0 & 0 & 0 & 0 & 0 & 0 & 0 & 0\\
\verb|algebra| & 44 & 0.0 & 6 & 0.0 & 0 & 0 & 0 & 0 & 0 & 0 & 0 & 0 & 0 & 0 & 0 & 17 & 17 & 17 & 17\\
\verb|vcg_simp| & 43 & 0.0 & 6 & 0.0 & 0 & 0 & 0 & 0 & 0 & 0 & 0 & 0 & 0 & 0 & 0 & 0 & 0 & 0 & 0\\
\verb|split| & 43 & 0.0 & 7 & 0.0 & 0 & 0 & 0 & 0 & 0 & 0 & 0 & 0 & 0 & 0 & 0 & 0 & 0 & 0 & 0\\
\verb|interfree_aux| & 39 & 0.0 & 2 & 0.0 & 0 & 0 & 0 & 0 & 0 & 0 & 0 & 0 & 0 & 0 & 0 & 0 & 0 & 0 & 0\\
\verb|hypsubst| & 37 & 0.0 & 7 & 0.0 & 0 & 0 & 0 & 0 & 0 & 0 & 0 & 0 & 0 & 0 & 0 & 0 & 0 & 0 & 0\\
\verb|sat| & 36 & 0.0 & 1 & 0.0 & 0 & 0 & 0 & 0 & 0 & 0 & 0 & 0 & 0 & 0 & 0 & 0 & 0 & 0 & 0\\
\verb|cooper| & 35 & 0.0 & 4 & 0.0 & 0 & 0 & 0 & 75 & 75 & 75 & 75 & 75 & 75 & 75 & 75 & 75 & 75 & 75 & 75\\
\verb|countable_datatype| & 32 & 0.0 & 1 & 0.0 & 100 & 100 & 100 & 100 & 100 & 100 & 100 & 100 & 100 & 100 & 100 & 100 & 100 & 100 & 100\\
\verb|simplesubst| & 29 & 0.0 & 4 & 0.0 & 0 & 0 & 0 & 0 & 0 & 0 & 0 & 0 & 0 & 0 & 0 & 0 & 0 & 25 & 25\\
\verb|vector| & 23 & 0.0 & 1 & 0.0 & 0 & 0 & 0 & 0 & 0 & 0 & 0 & 0 & 0 & 0 & 0 & 0 & 0 & 0 & 0\\
\verb|valid_certificate_tac| & 23 & 0.0 & 2 & 0.0 & 0 & 0 & 0 & 0 & 0 & 0 & 0 & 0 & 0 & 0 & 0 & 0 & 0 & 0 & 0\\
\verb|metis_exhaust| & 23 & 0.0 & 4 & 0.0 & 0 & 0 & 0 & 0 & 0 & 0 & 0 & 0 & 0 & 0 & 0 & 0 & 0 & 0 & 0\\
\verb|Seq_case_simp| & 23 & 0.0 & 3 & 0.0 & 0 & 0 & 0 & 0 & 0 & 0 & 67 & 67 & 100 & 100 & 100 & 100 & 100 & 100 & 100\\
\verb|oghoare| & 22 & 0.0 & 5 & 0.0 & 100 & 100 & 100 & 100 & 100 & 100 & 100 & 100 & 100 & 100 & 100 & 100 & 100 & 100 & 100\\
\verb|lifting| & 22 & 0.0 & 6 & 0.0 & 0 & 0 & 0 & 0 & 0 & 0 & 0 & 0 & 0 & 0 & 0 & 0 & 0 & 0 & 0\\
\verb|uint_arith| & 21 & 0.0 & 4 & 0.0 & 0 & 0 & 0 & 0 & 0 & 100 & 100 & 100 & 100 & 100 & 100 & 100 & 100 & 100 & 100\\
\verb|atomize| & 21 & 0.0 & 4 & 0.0 & 0 & 0 & 0 & 0 & 0 & 0 & 0 & 0 & 0 & 0 & 0 & 0 & 0 & 0 & 0\\
\verb|fresh_guess| & 19 & 0.0 & 2 & 0.0 & 0 & 0 & 0 & 0 & 0 & 0 & 0 & 0 & 0 & 0 & 0 & 0 & 0 & 0 & 0\\
\verb|unat_arith| & 18 & 0.0 & 4 & 0.0 & 0 & 0 & 0 & 0 & 0 & 0 & 0 & 25 & 75 & 75 & 75 & 75 & 75 & 75 & 75\\
\verb|hoare| & 17 & 0.0 & 2 & 0.0 & 50 & 50 & 50 & 50 & 50 & 50 & 50 & 50 & 50 & 50 & 50 & 50 & 100 & 100 & 100\\
\verb|approximation| & 17 & 0.0 & 2 & 0.0 & 0 & 0 & 0 & 0 & 0 & 0 & 0 & 0 & 0 & 0 & 0 & 0 & 0 & 0 & 0\\
\verb|annhoare| & 16 & 0.0 & 1 & 0.0 & 0 & 0 & 0 & 0 & 0 & 0 & 0 & 0 & 0 & 0 & 0 & 0 & 0 & 0 & 0\\
\verb|reify| & 15 & 0.0 & 1 & 0.0 & 0 & 0 & 0 & 0 & 0 & 0 & 0 & 0 & 0 & 0 & 0 & 0 & 0 & 0 & 0\\
\verb|record_auto| & 15 & 0.0 & 2 & 0.0 & 0 & 0 & 0 & 0 & 0 & 0 & 0 & 0 & 0 & 0 & 0 & 0 & 0 & 0 & 0\\
\verb|moura| & 15 & 0.0 & 4 & 0.0 & 0 & 0 & 0 & 0 & 0 & 0 & 0 & 0 & 0 & 0 & 0 & 0 & 0 & 0 & 0\\
\verb|intro_locales| & 15 & 0.0 & 2 & 0.0 & 0 & 0 & 0 & 0 & 0 & 0 & 50 & 50 & 50 & 50 & 50 & 50 & 50 & 50 & 50\\
\verb|bestsimp| & 15 & 0.0 & 3 & 0.0 & 0 & 0 & 0 & 0 & 0 & 0 & 0 & 0 & 0 & 0 & 0 & 0 & 0 & 0 & 0\\
\verb|induction_schema| & 14 & 0.0 & 1 & 0.0 & 0 & 0 & 0 & 0 & 0 & 0 & 0 & 0 & 0 & 0 & 0 & 0 & 0 & 0 & 0\\
\verb|pair_induct| & 13 & 0.0 & 2 & 0.0 & 0 & 0 & 100 & 100 & 100 & 100 & 100 & 100 & 100 & 100 & 100 & 100 & 100 & 100 & 100\\
\verb|finite_guess| & 9 & 0.0 & 1 & 0.0 & 0 & 0 & 0 & 0 & 0 & 0 & 0 & 0 & 0 & 0 & 0 & 0 & 0 & 0 & 0\\
\verb|merge_box| & 8 & 0.0 & 2 & 0.0 & 0 & 0 & 0 & 0 & 0 & 0 & 0 & 0 & 0 & 0 & 0 & 0 & 0 & 0 & 0\\
\verb|transfer_step| & 7 & 0.0 & 2 & 0.0 & 0 & 50 & 100 & 100 & 100 & 100 & 100 & 100 & 100 & 100 & 100 & 100 & 100 & 100 & 100\\
\verb|size_change| & 7 & 0.0 & 1 & 0.0 & 0 & 0 & 0 & 0 & 0 & 0 & 0 & 0 & 0 & 0 & 0 & 0 & 0 & 0 & 0\\
\verb|ns_induct| & 6 & 0.0 & 2 & 0.0 & 0 & 0 & 0 & 0 & 0 & 0 & 0 & 0 & 0 & 0 & 0 & 0 & 0 & 0 & 0\\
\verb|reflection| & 5 & 0.0 & 1 & 0.0 & 0 & 0 & 0 & 0 & 0 & 0 & 0 & 0 & 0 & 0 & 0 & 0 & 0 & 0 & 0\\
\verb|enabled| & 5 & 0.0 & 1 & 0.0 & 0 & 0 & 0 & 0 & 0 & 0 & 0 & 0 & 0 & 0 & 0 & 0 & 0 & 0 & 0\\
\verb|ml_tactic| & 4 & 0.0 & 1 & 0.0 & 0 & 0 & 0 & 0 & 0 & 0 & 0 & 0 & 0 & 0 & 0 & 0 & 0 & 0 & 0\\
\verb|mkex_induct| & 4 & 0.0 & 1 & 0.0 & 100 & 100 & 100 & 100 & 100 & 100 & 100 & 100 & 100 & 100 & 100 & 100 & 100 & 100 & 100\\
\verb|merge_temp_box| & 4 & 0.0 & 1 & 0.0 & 0 & 0 & 0 & 0 & 0 & 0 & 0 & 0 & 0 & 0 & 0 & 0 & 0 & 0 & 0\\
\verb|cartouche| & 3 & 0.0 & 1 & 0.0 & 0 & 0 & 0 & 0 & 0 & 0 & 0 & 0 & 0 & 0 & 0 & 0 & 0 & 0 & 0\\
\verb|analz_mono_contra| & 3 & 0.0 & 2 & 0.0 & 0 & 0 & 0 & 0 & 0 & 0 & 0 & 0 & 0 & 0 & 0 & 0 & 0 & 0 & 0\\
\verb|rename_client_map| & 2 & 0.0 & 1 & 0.0 & 0 & 0 & 0 & 0 & 0 & 0 & 0 & 0 & 0 & 0 & 0 & 0 & 0 & 0 & 0\\
\verb|possibility| & 2 & 0.0 & 1 & 0.0 & 0 & 0 & 0 & 0 & 0 & 0 & 0 & 0 & 0 & 0 & 0 & 0 & 0 & 0 & 0\\
\verb|defined| & 2 & 0.0 & 2 & 0.0 & 0 & 0 & 0 & 0 & 0 & 0 & 0 & 0 & 0 & 0 & 0 & 0 & 0 & 0 & 0\\
\verb|corec_unique| & 2 & 0.0 & 3 & 0.0 & 0 & 0 & 0 & 0 & 0 & 0 & 0 & 0 & 0 & 0 & 0 & 0 & 0 & 0 & 0\\
\verb|basic_possibility| & 2 & 0.0 & 2 & 0.0 & 0 & 0 & 0 & 0 & 0 & 0 & 0 & 0 & 0 & 0 & 0 & 0 & 0 & 0 & 0\\
\verb|sc_analz_freshK| & 1 & 0.0 & 1 & 0.0 & 0 & 0 & 0 & 0 & 0 & 0 & 0 & 0 & 0 & 0 & 0 & 0 & 0 & 0 & 0\\
\verb|invariant| & 1 & 0.0 & 1 & 0.0 & 0 & 0 & 0 & 0 & 0 & 0 & 0 & 0 & 0 & 0 & 0 & 0 & 0 & 0 & 0\\
\verb|disentangle| & 1 & 0.0 & 1 & 0.0 & 0 & 0 & 0 & 0 & 0 & 0 & 0 & 0 & 0 & 0 & 0 & 0 & 0 & 0 & 0\\
\verb|coherent| & 1 & 0.0 & 2 & 0.0 & 0 & 0 & 0 & 0 & 0 & 0 & 0 & 0 & 0 & 0 & 0 & 0 & 0 & 0 & 0\\
\verb|my_simp| & 0 & 0.0 & 1 & 0.0 & 0 & 0 & 0 & 0 & 0 & 0 & 0 & 0 & 0 & 0 & 0 & 0 & 0 & 0 & 0\\
\verb|dlo| & 0 & 0.0 & 1 & 0.0 & 0 & 0 & 0 & 0 & 0 & 0 & 0 & 0 & 0 & 0 & 0 & 0 & 0 & 0 & 0\\
\hline
\end{tabular}
\end{center}
\end{table*}

\begin{table*}[ht]
\caption{Evaluation of \pamper{} on proof methods defined by users.}
\label{table:user}
\begin{center}
\begin{tabular}{l r r r r r r r r r r r r r r r r r r r r r r}
\hline\noalign{\smallskip}
proof method & training & \% & evaluation & \% & 1 & 2 & 3 & 4 & 5 & 6 & 7 & 8 & 9 & 10 & 11 & 12 & 13 & 14 & 15\\
\hline
\noalign{\smallskip}
\verb|refine_vcg| & 193 & 0.1 & 31 & 0.1 & 0 & 0 & 0 & 36 & 36 & 36 & 36 & 36 & 36 & 36 & 36 & 36 & 36 & 36 & 36\\
\verb|inv_cterms| & 173 & 0.0 & 17 & 0.0 & 0 & 0 & 0 & 0 & 0 & 24 & 71 & 71 & 100 & 100 & 100 & 100 & 100 & 100 & 100\\
\verb|autoref_monadic| & 167 & 0.0 & 12 & 0.0 & 0 & 0 & 58 & 58 & 58 & 58 & 58 & 58 & 58 & 58 & 58 & 67 & 75 & 75 & 75\\
\verb|autoref| & 113 & 0.0 & 17 & 0.0 & 0 & 0 & 0 & 0 & 0 & 12 & 35 & 59 & 100 & 100 & 100 & 100 & 100 & 100 & 100\\
\verb|sep_auto| & 108 & 0.0 & 6 & 0.0 & 0 & 0 & 0 & 67 & 83 & 83 & 100 & 100 & 100 & 100 & 100 & 100 & 100 & 100 & 100\\
\verb|sepref| & 86 & 0.0 & 12 & 0.0 & 0 & 0 & 0 & 100 & 100 & 100 & 100 & 100 & 100 & 100 & 100 & 100 & 100 & 100 & 100\\
\verb|hoare_rule| & 80 & 0.0 & 7 & 0.0 & 0 & 14 & 14 & 14 & 14 & 14 & 14 & 14 & 14 & 14 & 14 & 14 & 14 & 14 & 14\\
\verb|refine_rcg| & 77 & 0.0 & 10 & 0.0 & 0 & 0 & 0 & 0 & 0 & 0 & 0 & 0 & 0 & 0 & 10 & 30 & 50 & 50 & 50\\
\verb|solves| & 73 & 0.0 & 8 & 0.0 & 0 & 0 & 0 & 0 & 0 & 0 & 0 & 0 & 0 & 0 & 0 & 38 & 38 & 38 & 38\\
\verb|prec| & 70 & 0.0 & 7 & 0.0 & 0 & 0 & 0 & 0 & 0 & 0 & 0 & 0 & 0 & 0 & 0 & 0 & 0 & 0 & 0\\
\verb|vcg_step| & 66 & 0.0 & 5 & 0.0 & 0 & 0 & 0 & 0 & 40 & 80 & 80 & 80 & 80 & 80 & 80 & 80 & 80 & 80 & 80\\
\verb|vc_solve| & 58 & 0.0 & 5 & 0.0 & 0 & 0 & 0 & 0 & 0 & 0 & 0 & 0 & 0 & 0 & 0 & 0 & 0 & 0 & 0\\
\verb|code_simp| & 42 & 0.0 & 2 & 0.0 & 0 & 0 & 0 & 0 & 0 & 0 & 0 & 0 & 0 & 0 & 0 & 0 & 0 & 0 & 0\\
\verb|refine_dref_type| & 41 & 0.0 & 1 & 0.0 & 0 & 0 & 0 & 0 & 0 & 0 & 0 & 0 & 0 & 100 & 100 & 100 & 100 & 100 & 100\\
\verb|vcg_jackhammer| & 40 & 0.0 & 3 & 0.0 & 0 & 0 & 0 & 0 & 0 & 0 & 0 & 0 & 0 & 0 & 0 & 0 & 0 & 0 & 0\\
\verb|refine_transfer| & 35 & 0.0 & 5 & 0.0 & 0 & 0 & 0 & 0 & 0 & 0 & 0 & 0 & 0 & 20 & 20 & 40 & 40 & 40 & 40\\
\verb|parametricity| & 35 & 0.0 & 3 & 0.0 & 0 & 0 & 0 & 0 & 0 & 0 & 0 & 0 & 0 & 0 & 0 & 0 & 0 & 0 & 0\\
\verb|sepref_to_hoare| & 33 & 0.0 & 6 & 0.0 & 0 & 0 & 0 & 0 & 0 & 0 & 0 & 0 & 0 & 0 & 0 & 0 & 0 & 0 & 0\\
\verb|separata| & 31 & 0.0 & 5 & 0.0 & 0 & 0 & 0 & 60 & 60 & 60 & 60 & 60 & 60 & 60 & 60 & 60 & 60 & 60 & 60\\
\verb|vcg_ni| & 29 & 0.0 & 1 & 0.0 & 0 & 0 & 0 & 0 & 0 & 0 & 0 & 0 & 0 & 0 & 0 & 0 & 0 & 0 & 0\\
\verb|sepref_dbg_trans_step| & 21 & 0.0 & 4 & 0.0 & 0 & 0 & 0 & 0 & 0 & 0 & 50 & 75 & 75 & 75 & 75 & 75 & 75 & 75 & 75\\
\verb|clarsimp_all| & 19 & 0.0 & 2 & 0.0 & 0 & 0 & 0 & 0 & 0 & 0 & 0 & 0 & 0 & 0 & 0 & 0 & 0 & 0 & 0\\
\verb|vcg_nihe| & 18 & 0.0 & 4 & 0.0 & 0 & 0 & 0 & 0 & 0 & 0 & 0 & 0 & 0 & 0 & 0 & 0 & 0 & 0 & 0\\
\verb|tagged_solver| & 11 & 0.0 & 2 & 0.0 & 0 & 0 & 0 & 0 & 0 & 0 & 0 & 0 & 0 & 0 & 0 & 0 & 0 & 0 & 0\\
\verb|rprems| & 10 & 0.0 & 1 & 0.0 & 0 & 0 & 0 & 0 & 0 & 0 & 0 & 0 & 0 & 0 & 0 & 0 & 0 & 0 & 0\\
\verb|akra_bazzi_termination| & 10 & 0.0 & 2 & 0.0 & 0 & 0 & 100 & 100 & 100 & 100 & 100 & 100 & 100 & 100 & 100 & 100 & 100 & 100 & 100\\
\verb|kat_hom| & 8 & 0.0 & 1 & 0.0 & 0 & 0 & 0 & 0 & 0 & 0 & 0 & 0 & 0 & 0 & 0 & 0 & 0 & 0 & 0\\
\verb|applicative_nf| & 8 & 0.0 & 3 & 0.0 & 0 & 0 & 0 & 0 & 0 & 0 & 0 & 0 & 0 & 0 & 0 & 0 & 0 & 0 & 0\\
\verb|wellformed| & 6 & 0.0 & 2 & 0.0 & 0 & 0 & 0 & 0 & 0 & 0 & 0 & 0 & 0 & 0 & 0 & 0 & 0 & 0 & 0\\
\verb|seq_stop_inv_method| & 5 & 0.0 & 1 & 0.0 & 0 & 0 & 0 & 0 & 0 & 0 & 100 & 100 & 100 & 100 & 100 & 100 & 100 & 100 & 100\\
\verb|fo_rule| & 5 & 0.0 & 1 & 0.0 & 0 & 0 & 0 & 0 & 0 & 0 & 0 & 0 & 0 & 0 & 0 & 0 & 0 & 0 & 0\\
\verb|dlo_reify| & 5 & 0.0 & 1 & 0.0 & 0 & 0 & 0 & 0 & 0 & 0 & 0 & 0 & 0 & 0 & 0 & 0 & 0 & 0 & 0\\
\verb|refine_mono| & 4 & 0.0 & 1 & 0.0 & 0 & 0 & 0 & 0 & 0 & 0 & 0 & 0 & 0 & 0 & 0 & 0 & 0 & 0 & 0\\
\verb|sepref_dbg_opt_init| & 3 & 0.0 & 1 & 0.0 & 0 & 0 & 0 & 0 & 0 & 0 & 0 & 0 & 0 & 0 & 0 & 0 & 0 & 0 & 0\\
\verb|sepref_dbg_cons_init| & 3 & 0.0 & 1 & 0.0 & 0 & 0 & 0 & 0 & 0 & 0 & 0 & 0 & 0 & 0 & 0 & 0 & 0 & 0 & 0\\
\verb|sep_subst| & 3 & 0.0 & 1 & 0.0 & 0 & 0 & 0 & 0 & 0 & 0 & 0 & 0 & 0 & 0 & 0 & 0 & 0 & 0 & 0\\
\verb|vcg_jackhammer_ff| & 2 & 0.0 & 1 & 0.0 & 0 & 0 & 0 & 0 & 0 & 0 & 0 & 0 & 0 & 0 & 0 & 0 & 0 & 0 & 0\\
\verb|sep_select| & 2 & 0.0 & 1 & 0.0 & 0 & 0 & 0 & 0 & 0 & 0 & 0 & 0 & 0 & 0 & 0 & 0 & 0 & 0 & 0\\
\hline
\end{tabular}
\end{center}
\end{table*}

\newpage

\begin{figure*}[!ht]
      \centerline{\includegraphics[width=160mm]{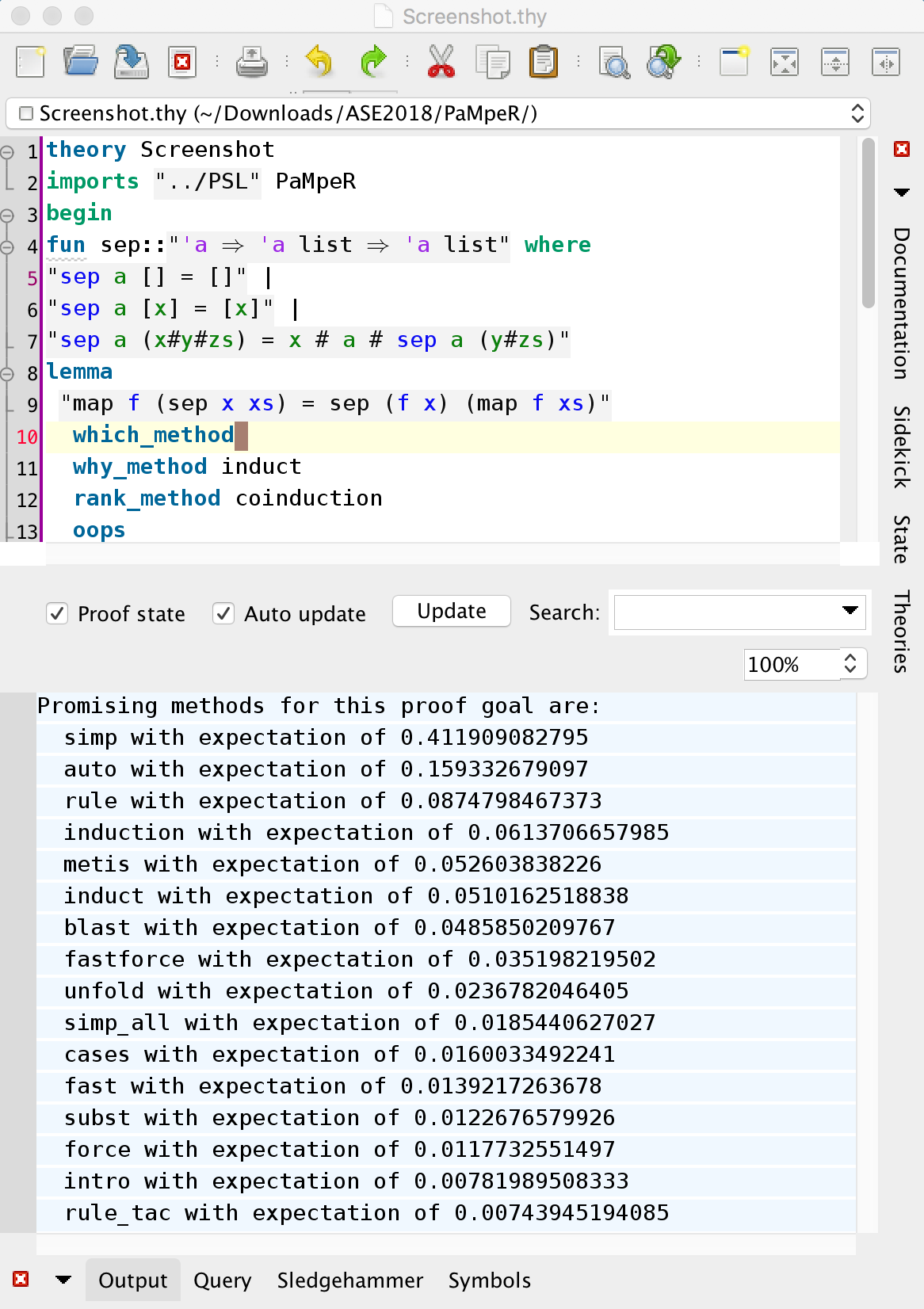}}
      \captionsetup{labelformat=empty}
      \caption{Screenshot of Isabelle/HOL with \pamper{}.}
      \label{fig:screenshot}
\end{figure*}

\end{document}